\documentclass[manuscript]{aastex62}

\usepackage{amsmath} 

\usepackage{graphicx}
\usepackage[caption=false]{subfig}

\usepackage{rotating} 

\graphicspath{ {/users/aarnold/Research/PaperDocuments/paper_1/Overleaf_Hard_Copy/KELT_Mira_Catalog/} }

\received{Month Day, 2018}
\revised{Month Day 2018}
\accepted{\today}
\submitjournal{ApJ}

\shorttitle{Long-Period High-Amplitude Red Variables in the KELT Survey}
\shortauthors{Arnold, McSwain, Pepper et.\ al.}

\begin{document}
\title{Long-Period High-Amplitude Red Variables in the KELT Survey}

\author{R. Alex Arnold}
\affiliation{Lehigh University, Department of Physics, 16 Memorial Drive East, Bethlehem, PA, 18015, USA}

\author[0000-0002-4775-2803]{M. Virginia McSwain}
\affiliation{Lehigh University, Department of Physics, 16 Memorial Drive East, Bethlehem, PA, 18015, USA}

\author[0000-0002-3827-8417]{Joshua Pepper}
\affiliation{Lehigh University, Department of Physics, 16 Memorial Drive East, Bethlehem, PA, 18015, USA}

\author[0000-0002-4678-4432]{Patricia A. Whitelock}
\affiliation{South African Astronomical Observatory, P.O. Box 9, Observatory, 7935 Cape Town, South Africa}
\affiliation{Department of Astronomy, University of Cape Town, 7701 Rondebosch, South Africa}

\author[0000-0003-1681-0430]{Nina Hernitschek}
\affiliation{Department of Physics \& Astronomy, Vanderbilt University, 6301 Stevenson Center Lane, Nashville, TN 37235, USA} 
\affiliation{Data Science Institute, Vanderbilt University, 2414 Highland Avenue Nashville, TN 37212, USA}

\author[0000-0001-5160-4486]{David J. James}
\affiliation{Center for Astrophysics $\vert$ Harvard \& Smithsonian, 60 Garden Street, Cambridge, MA 02138, USA}
\affiliation{Black Hole Initiative at Harvard University, 20 Garden Street, Cambridge, MA 02138, USA}

\author[0000-0002-4236-9020]{Rudolf B. Kuhn}
\affiliation{South African Astronomical Observatory, P.O. Box 9, Observatory, 7935 Cape Town, South Africa}
\affiliation{Southern African Large Telescope, P.O. Box 9, Observatory, 7935, Cape Town, South Africa}

\author[0000-0002-1825-0097]{Michael B. Lund}
\affiliation{Caltech/IPAC-NExScI, Pasadena, CA 91125 USA}

\author[0000-0001-8812-0565]{Joseph E. Rodriguez} 
\affiliation{Center for Astrophysics $\vert$ Harvard \& Smithsonian, 60 Garden Street, Cambridge, MA 02138, USA}

\author[0000-0001-5016-3359]{Robert J. Siverd}
\affiliation{Department of Physics \& Astronomy, Vanderbilt University, 6301 Stevenson Center Lane, Nashville, TN 37235, USA}

\author[0000-0002-3481-9052]{Keivan G. Stassun}
\affiliation{Department of Physics \& Astronomy, Vanderbilt University, 6301 Stevenson Center Lane, Nashville, TN 37235, USA}

\begin{abstract}
 We present a sample of 4,132 Mira-like variables (red variables with long periods and high amplitudes) in the Kilodegree Extremely Little Telescope (KELT) survey. Of these, 814 are new detections. We used 2MASS colors to identify candidate asymptotic giant branch (AGB) stars. We tested for photometric variability among the sample and used Lomb-Scargle to determine the periodicity of the variable sample. We selected variables with high amplitudes and strong periodic behavior using a Random Forest classifier. Of the sample of 4,132 Mira-like variables, we estimate that 70\% are Miras, and 30\% are semi-regular (SR) variables. We also adopt the method of using ($W_{RP} - W_{K_s}$) vs.\ ($J - K_s$) colors \citep{lebzelter18} in distinguishing between O-rich and C-rich Miras and find it to be an improvement over 2MASS colors.

\end{abstract}

\section{Introduction}\label{sec:intro}

Late-type, low- to intermediate-mass red giant stars are almost always unstable against pulsation and form a class of variable stars known as the long-period variables (LPVs). Of these, Miras are perhaps the best known. Miras are a short-lived phase of a low- to intermediate-mass star's life when it has evolved to the tip of the asymptotic giant branch (AGB). At this stage, these stars exhibit large bolometric luminosities and this, combined with their low average densities and high-amplitude pulsations, causes them to experience significant mass loss through stellar winds before they meet their eventual fate as white dwarfs surrounded by transient planetary nebulae. This process enriches the interstellar medium with heavy elements, in particular carbon and \textit{s}-process elements, and therefore the chemistry of Miras directly influences the composition of stars and the overall chemical evolution of galaxies. Aside from their importance to Galactic chemistry, Miras also show promise as distance indicators. Multiple studies (\citealp{feast89}; \citealp{hughes90}; \citealp{whitelock08}; and for a general overview see \citealp{whitelock12} and references therein) have found evidence for a period-luminosity (PL) relation in the near infrared for Miras. In the future they may offer a viable alternative to Cepheids as extragalactic standard candles \citep{feast02, whitelock13, huang18}. 

Miras are conventionally defined to have periods \textgreater{} 100 days and amplitudes of $\Delta V > 2.5$ mag \citep{payne51, samus17}, but their amplitudes decrease at longer wavelengths with typical amplitudes of $\Delta I > 0.8$ \citep{soszynski09} and $\Delta K > 0.4$ mag \citep{whitelock06}. While an amplitude criterion can effectively identify Miras, the specific amplitude limits are somewhat arbitrary \citep{whitelock96a}.

Another class of LPVs are the semi-regular variables (SRs). They are less studied than Miras and are defined to have amplitudes of $\Delta V < 2.5$ \citep{payne51, samus17}. SR variables represent a more heterogenous class than Miras and consist of stars on both the AGB and red giant branch (RGB) with a range of masses \citep{whitelock96b}. Their light curves typically exhibit more irregular behavior than Miras, with some exhibiting signs of multi-periodicity \citep{cadmus15}. Despite what their name implies, SRs can exhibit regular periodic behavior similar to Miras but with smaller amplitudes. This distinguishing amplitude criterion between Miras and SRs is arbitrary, and it is possible that low-amplitude Miras and high-amplitude SRs may share physical similarities \citep{kerschbaum92, bedding98}. We should note that even the condition of \textit{regular} periodic behavior is arbitrary since Miras themselves are not necessarily regular pulsators and can exhibit cycle-to-cycle amplitude changes \citep{whitelock96b}. In particular, carbon-rich Mira variables experience asymmetric mass loss which causes irregularities in their light curves (e.g. see the visual light curve of R For in Figure 6 of \citet{whitelock97}). Observations irregularly spaced in time can also obscure the regularity of Mira and SR light curves, making such a distinction difficult. For these reasons, we will attempt to discuss high-amplitude, long-period red variable stars without discriminating between Miras and SRs with near-regular periodicity. We will refer to these stars of interest as \textit{Mira-like} variables.

The cause of variability in Miras and SRs is radial pulsation which occurs in either fundamental or overtone modes. An important discovery by \citet{wood99} showed that LPVs, which include Miras and SRs, populate multiple PL sequences. Recent studies (\citealp{trabucchi17} and references therein) have determined that these sequences are due to the dominant pulsating mode present in the star, with Miras corresponding to fundamental mode pulsation. Most SRs are overtone pulsators, but \citet{bedding98} found a majority of their SR sample fell on the same PL sequence as Miras. These studies suggest a more physically meaningful distinction between classes of LPVs would be the dominant mode of the star's pulsation instead of an amplitude cut-off, but mode determination is difficult without information on both their periods and luminosities. We do not attempt to pursue this method of classification in this work, but it may be worth future analysis. Sources with luminosities determined in future work can be used for refining PL relations, and comparisons between observed and modeled pulsation modes can serve as tests of stellar interior models.

Because of their high luminosity and scientific value, many catalogs of LPVs have been produced by sky surveys such as those from the MACHO project \citep{alcock93, wood99}, the All-Sky Automated Survey (ASAS) \citep{pojmanski02, vogt16}, the Northern Sky Variability Survey (NSVS) \citep{wozniak04a, wozniak04b}, the Optical Gravitational Lensing Experiment (OGLE) \citep{udalski92, soszynski09, soszynski11, soszynski13}, and most recently \textit{Gaia} \citep{gaia16, gaia18, mowlavi18}. KELT provides advantages over several previous surveys due to its high photometric precision, long time baseline of observations, and its coverage of both the northern and southern sky. It also detects bright stars which make for easier follow-up observations.

We have created a catalog of 4,132 Mira-like variables from the KELT survey. These are high-amplitude, long-period red variables which are predominantly Miras. Section 2 of our paper discusses the KELT data and the cross-match between KELT and the 2MASS survey. Our method for creating our catalog is presented in section 3. In section 4 we discuss results of our catalog and compare it to similar catalogs such as that of \citet{vogt16}. Section 5 discusses a technique of classifying Miras as oxygen- or carbon-rich based on the analysis of \citet{lebzelter18}. We summarize these results in section 6.

\section{Observations and Data Reduction}

\subsection{Observations}\label{sec:obs}

The observational data used in this study are time-series photometry taken from the KELT survey \citep{pepper07, pepper12}. KELT is a photometric survey consisting of two small-aperture (42 mm) wide-field ($26^{\circ}\times26^{\circ}$) telescopes originally designed to search for transiting planets orbiting bright stars with 8 \textless{} \textit{V} \textless{} 13 and began operations in 2005. The KELT survey uses an effective passband roughly equivalent to a very broad $R$-band filter. The survey performs observations in both the northern and southern sky, with the northern telescope at Winer Observatory in Arizona, United States, and the southern telescope at the South African Astronomical Observatory near Sutherland, South Africa. With these two locations, KELT observes over 70\% of the sky. The time baselines of KELT observations vary, with some of the longest up to 10 years. These long baselines, combined with a typical cadence of 30 minutes and high photometric precision, result in the KELT survey being a valuable tool for studying phenomena beyond exoplanet detection.

\begin{figure}[h]
\centering 
\includegraphics[scale = 1]{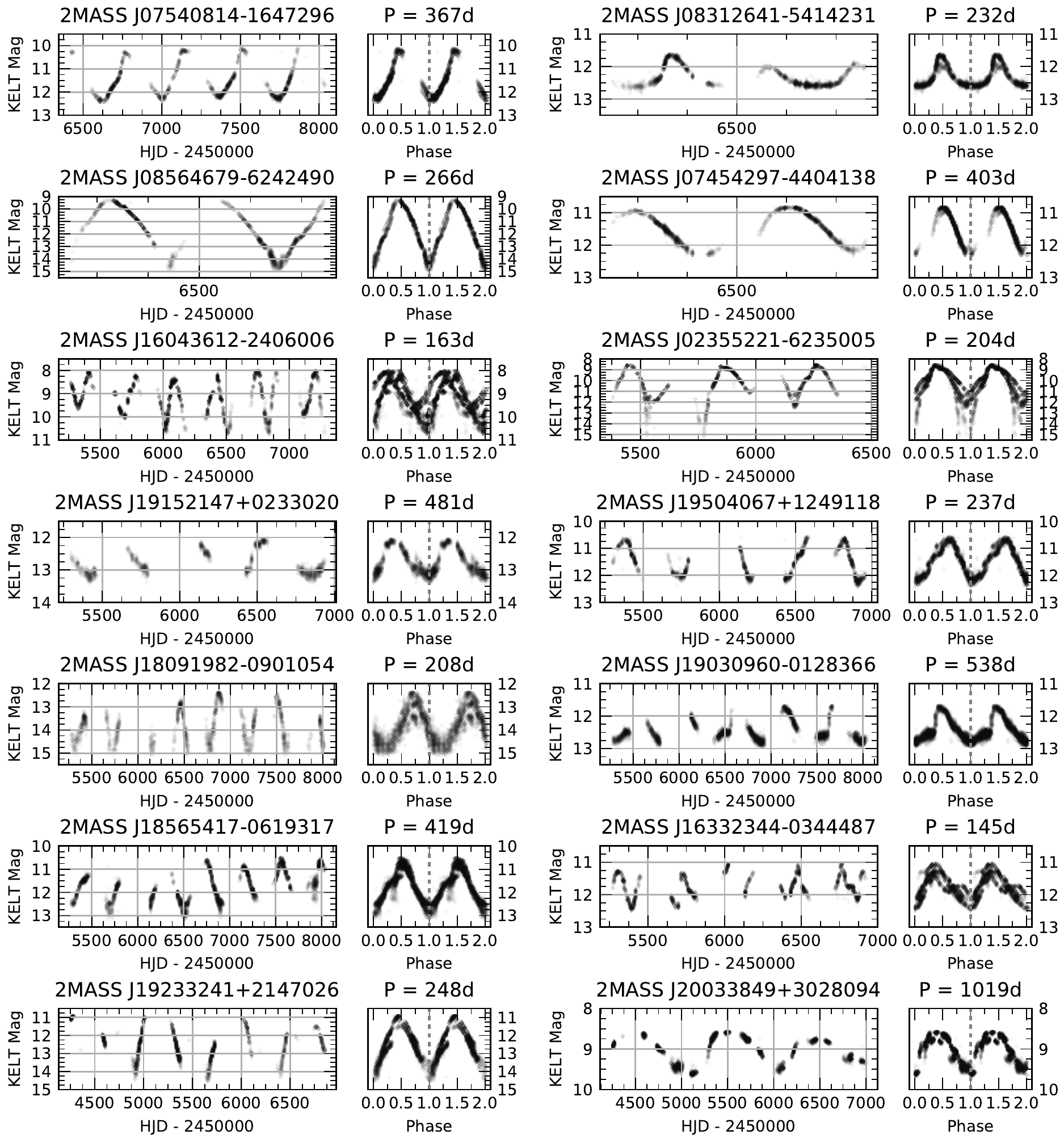}
\caption{A sample of 14 light curves from our Mira-like catalog spanning various periods and amplitudes. Two cases from this sample illustrate data reduction challenges discussed in the text. Star 2MASS J08312641-5414231 has flattened minima, and thus a reduced amplitude, due to blending. Star 2MASS J02355221-6235005 exhibits noisy minima in its eastern data which can occur as the light curve approaches KELT's faint detection limit under bright sky conditions.}
\label{fig:sample_lcs}
\end{figure}

One point of consideration is the large pixel scale of the KELT survey ($\sim$23 arcsec), which can result in stars in the KELT image blending together. This blending between a target star and nearby neighbors can cause two possible effects: variability from the neighboring sources appearing in the light curve of a target, and contaminating sources diluting the variability of the target star. This latter effect dampens the amplitude of the magnitude variations and flattens the light curve minima (see 2MASS J08312641-5414231 in Figure \ref{fig:sample_lcs}). Because the variables we are interested in are typically brighter than nearby neighbors, and the high-amplitude long-period variability of Miras is not typically seen in other variable stars, we assume the variability is attributed to the target star in almost all cases. Due to the high amplitudes of Miras, any dilution must be significant to affect our ability to detect Miras. We discuss attempts to identify and correct for blending in \S \ref{sec:rfresults}.

When stars fade to a brightness below the background sky, our ability to obtain reliable differential photometry becomes dominated by systematic noise, and the pipeline-generated light curves display unphysically deep features.  Since the point where that occurs depends on the sky background, degree of nearby star blending, PSF size and shape, and pixel quantum efficiency, we have been unable to implement an automated cut to remove such behavior.  Some of the Miras in our sample show this effect at their photometric minima (such as 2MASS J02355221-6235005 in Figure \ref{fig:sample_lcs}), but we have not seen any cases where this effect causes spurious detections in our sample.

\begin{figure}[t]
\centering 
\includegraphics[scale = .5]{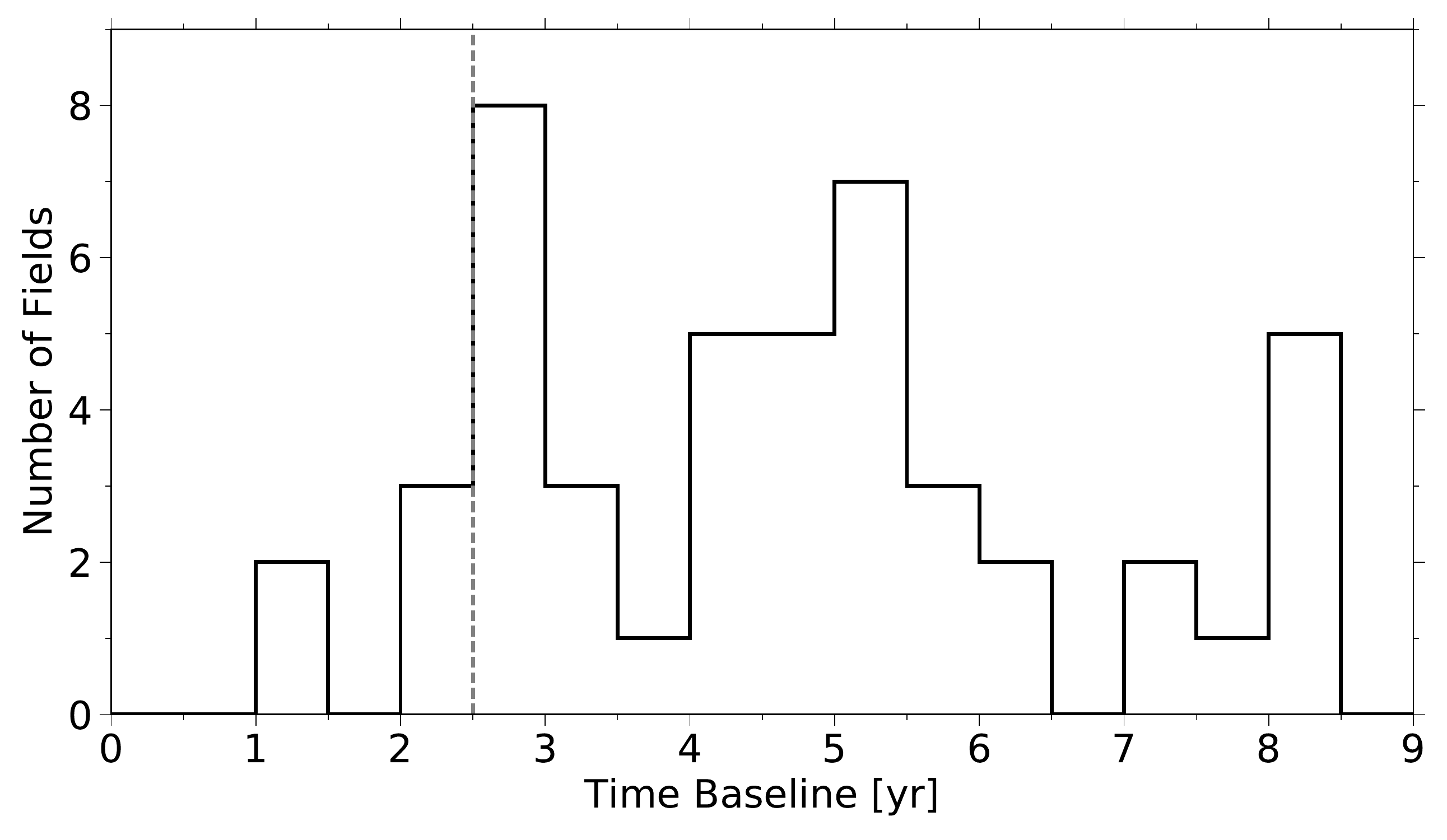}
\caption{The maximum time baseline of the various KELT fields used in this study. KELT consists of older fields nearing almost 10 years of observations and newer fields with baselines between 1-4 years. For our study we removed fields with baselines less than 2.5 years. This has been indicated by the grey dashed line.}
\label{fig:baseline}
\end{figure}

Due to the German Equatorial Mount used by KELT, photometric data for a target star are observed in separate eastern and western orientations. The data for each orientation are reduced separately resulting in most KELT objects having two light curves, but there do exist KELT objects only identified in one orientation. When necessary, we have created combined light curves for the KELT objects by combining their eastern and western data. A magnitude offset typically exists between the two orientations that must be determined for each object before creating the combined light curve. We calculated this magnitude offset for each KELT object by first dividing the eastern and western light curves into identical one day bins. We determined the median magnitude of each bin and then calculated the differences of the eastern and western medians for each day. The median of these differences is the eastern/western offset. After applying the offset correction for each KELT object, the eastern and western light curves were combined into one light curve. A photometric error of 0.1 mag was assumed for all photometric data.

The KELT data are divided into fields based on sky location. There are 23 fields observed by the northern telescope and 24 by the southern telescope. Fields that have been in operation earlier, e.g.\ N01, have baselines nearing 10 years, while fields that have begun operations more recently, e.g.\ N16, have shorter time baselines between 1-4 years.  Because Miras have long periods, typically 100-1,000 days, we removed from our study KELT fields with baselines of observations less than 2.5 years from our study (Figure \ref{fig:baseline}).

\subsection{Cross-Matching KELT to 2MASS}\label{sec:crossmatch}

To determine the 2MASS photometry of each source, we cross-matched the list of KELT stars to the 2MASS catalog \citep{skruskie06} using a 25 arcsec search radius and considered only 2MASS objects with $J$ magnitude brighter than 16. We chose this search radius as it is approximately the size of a KELT pixel, and smaller radii were found to exclude potential matches. This radius can result in multiple matches to 2MASS; therefore, we compared the KELT magnitude, $R_{KELT}$, to $J$ for each 2MASS object in the search radius. We are interested in red variables, and therefore excluded 2MASS objects with ($R_{KELT} - J$) \textless{} $-1$. The 2MASS object with the brightest $J$ magnitude was then chosen as the appropriate match. These search criteria were able to match 97\% of the KELT catalog to a 2MASS object. Note that for the remainder of the paper we refer to the catalog of KELT objects which match to 2MASS as the KELT catalog.

\begin{figure}[t]
\centering 
\includegraphics[scale = .5]{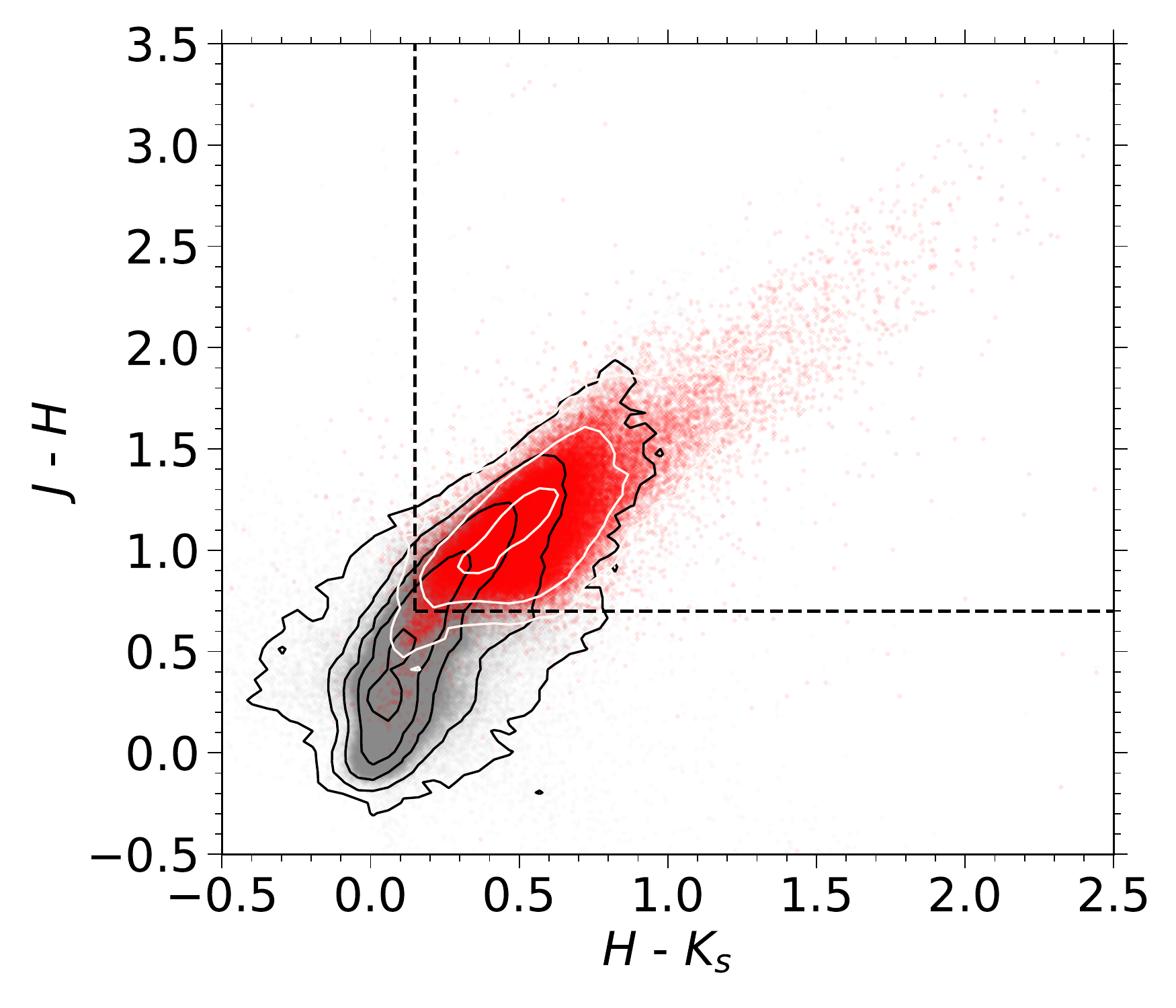}
\caption{Color cuts of the KELT sample (grey) and the VSX Miras and semi-regulars (red). Our color cuts of ($H - K_s$) \textgreater{} 0.15 and ($J - H$) \textgreater{} 0.7 are shown as dashed lines. Contours express the density of plotted data. The $H - K_s$ and $J - H$ data are divided into 0.05 mag bins and the contours are logarithmically spaced to circumscribe bins containing 100 to 100,000 data points.}
\label{fig:color_cuts}
\end{figure}

\section{Methods}

\subsection{Preliminary Cuts for Selecting Variable Candidates}\label{sec:cuts}

To identify candidate AGB stars, we used a color-color diagram of ($H - K_s$) vs.\ ($J - H$) \citep{feast82}. To justify our color cut criteria, we first selected a sample of known Miras and SRs from the International Variable Star Index (VSX) from the American Association of Variable Star Observers (AAVSO) (\citealp{watson06} using the 2019 version of the database). We then matched these to the KELT catalog using the matched 2MASS coordinates. We used the 2MASS colors of this matched VSX sample and inspected their color-color distribution to determine appropriate cuts. Figure \ref{fig:color_cuts} shows the ($H - K_s$) vs.\ ($J - H$) colors of all KELT matches in grey and VSX Miras and SRs in red. We selected stars with ($H - K_s$) \textgreater{} 0.15 and ($J - H$) \textgreater{} 0.7 as candidate AGB stars. These cuts are shown by the dashed black lines in Figure \ref{fig:color_cuts}.

\begin{figure}[t]
\centering 
\includegraphics[scale = .5]{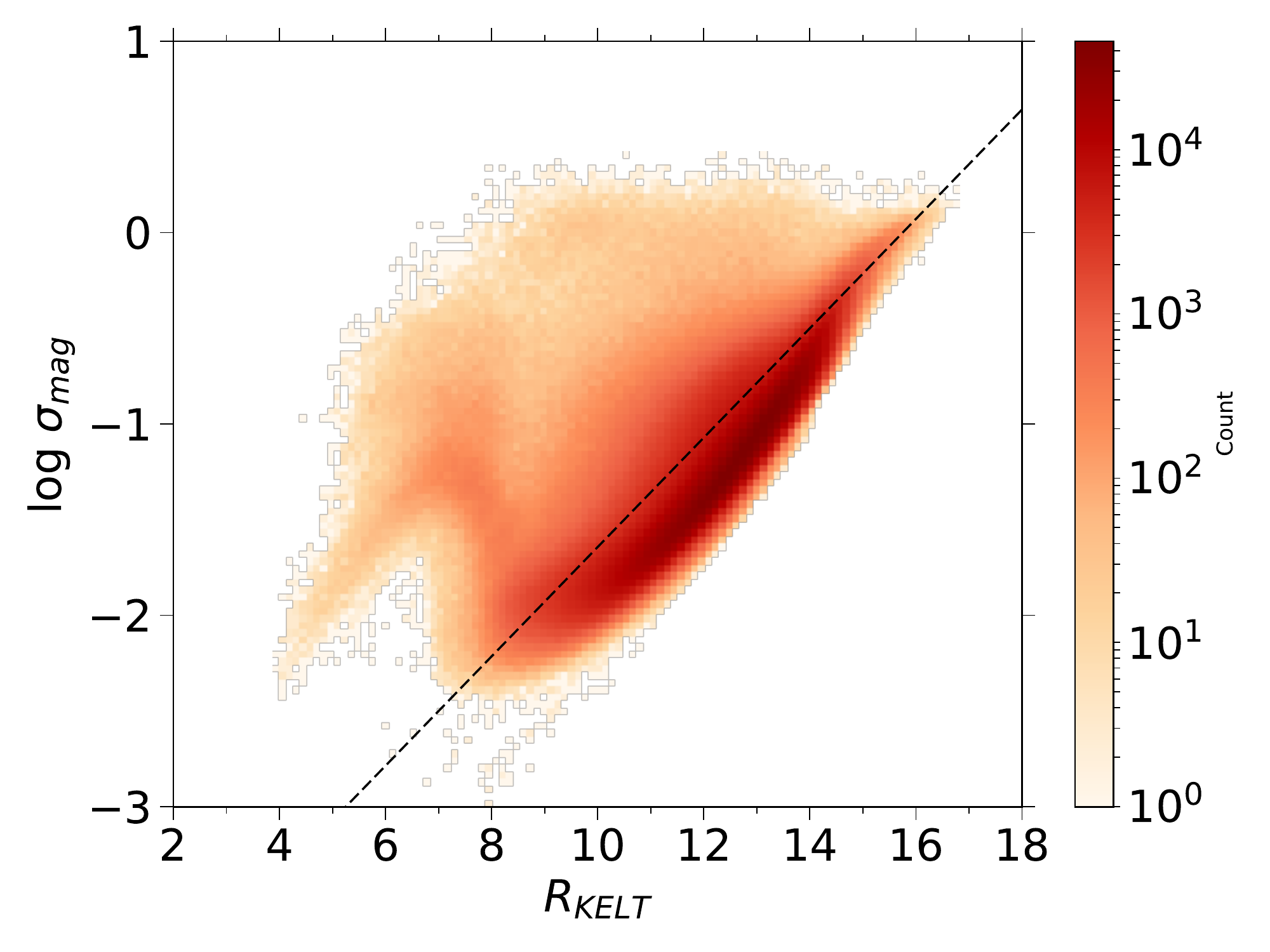}
\caption{A two-dimensional histogram of the $R_{KELT}$ and standard deviation of magnitudes of the eastern and western KELT light curves in the color cut sample. Objects with either an eastern or western light curve above the dashed line were selected as candidate variables and kept for further study. The increase in standard deviations at brighter magnitudes is due to saturation effects.}
\label{fig:rms}
\end{figure}

After the candidate AGB stars were found, we searched for variability among these objects. We first identified candidate AGB objects with potential variability based on the light curve mean magnitude and standard deviation of the magnitudes. This allows us to quickly select for probable variable stars. We desired to reduce the number of objects under consideration before combining light curves. Therefore, we calculated the standard deviation of the eastern and western light curves magnitudes separately. Individually, an eastern or western light curve may not cover a full phase of variability; thus, we classified an object as a candidate variable if the standard deviation of either light curve lies above the line shown in Figure \ref{fig:rms}.

After the standard deviation cuts, we found cases where the standard deviation was inflated by outlying data points in the light curve. To account for these cases, we rejected stars with eastern and western light curve amplitudes less than 0.2 mag after removing the 5\% and 95\% outlying data points. For the remainder of our analysis we removed only the 1\% and 99\% outlying data points so as to not remove data for the later period analysis. After applying these amplitude cuts, we also removed light curve data points with KELT magnitude fainter than 15 in order to remove noisy data likely below the faint detection limit of KELT. This magnitude cut is effective in removing spurious data, but we note there still exist a sample of light curves exhibiting noisy minima. We then combined eastern and western light curves using the method described in \S \ref{sec:obs}. Collectively, these cuts primarily remove non-red, non-variable objects, and allow us to greatly reduce the light curve data considered for further analysis by 98\% (from 4.2 million KELT objects to 70,000).

\subsection{Selecting Mira-Like Variables with Random Forest}\label{sec:rfmain}

We applied a Random Forest (RF) classifier \citep{breiman01} to identify Mira-like variables among objects passing the aforementioned cuts. RF is an ensemble machine learning method which uses multiple de-correlated decision trees to predict the classification type of an object from various measurable properties (features) of the object. The decision trees ``learn" the classification types using a training set of data containing features of objects with known classifications (labels). To de-correlate the decision trees, each is built on a random selection of the training data, and a random subset of the total features are used at each decision node. Random Forest has proven effective in classifying variable stars \citep{richards11, dubath11, hernitschek16, pashch18}, and in cases has been shown to outperform other machine learning classifiers such as support vector machines \citep{yuan17}. 

The RF classifier requires features for each star which ideally characterize the object. The classifier also needs a training set of stars with features and known labels (e.g. Mira-like or not). Once the RF classifier is trained it can be used to classify a set of stars with unknown classifications. We used \textsc{Scikit-Learn} \citep{pedregosa12}, an open-source machine learning library for \textsc{Python}, to build our RF classifier. For each star passing the cuts in \S \ref{sec:cuts}, we determined 16 features. Of these, 5 are determined from the combined light curve data, 3 from a period search, and 8 are photometric properties from KELT and 2MASS. All features are listed in Table 1 along with the source of the feature (light curve, periodogram, photometry) and the rank of the feature's importance in the RF classification as determined by \textsc{Scikit-Learn}. These features are discussed in the following subsections. 

\subsubsection{Features from Photometry}\label{sec:rfphotometry}

The features from KELT and 2MASS photometry are $R_{KELT}$, $J$, $H$, $K_s$, ($R_{KELT} - J$), ($J - H$), ($J - K_s$), and ($H - K_s$). From \S \ref{sec:cuts} we found most Miras and other red variables have ($H - K_s) > 0.15$, and we found that Miras in our VSX sample are shifted toward redder $H - K_s$ colors compared to non-Mira red variables. We therefore retained this color as a RF feature and kept the remaining colors and magnitudes as RF features for comparison. Consulting Table 1, though, we see that, aside from $H - K_s$, these photometric features have the least importance in classifying our Mira-like objects.

\begin{table}[h!]
  \begin{center}
    \caption{Features for the Random Forest Classifier}
    \label{tab:table1}
    \begin{tabular}{l c r}
     \hline \hline
      Feature & Source & Rank\\
      \hline
      $\sigma_{mag}$  & light curve & 1\\
      $A_{10,90}$  & light curve & 2\\
      $A_{1,99}$   & light curve & 3\\
      Welch-Stetson $L$   & light curve & 4\\
      LS Power   & periodogram & 5\\
      Alarm & light curve & 6\\
      $H - K_s$ & photometry & 7\\
      LS Period & periodogram & 8\\
      LS S/N & periodogram & 9\\
      $R_{KELT}$ & photometry & 10\\
      $R_{KELT} - J$ & photometry & 11\\
      $J - H$ & photometry & 12\\
      $K_s$ & photometry & 13\\
      $J - K_s$ & photometry & 14\\
      $J$ & photometry & 15\\
      $H$ & photometry & 16\\      
    \end{tabular}
  \end{center}
\end{table}

\subsubsection{Features from the Light Curve}\label{sec:rflightcurve}

From each combined light curve we determined the standard deviation of magnitudes $\sigma_{mag}$, 10th-90th percentile amplitude $A_{10,90}$, 1st-99th percentile amplitude $A_{1,99}$, Welch-Stetson $L$ \citep{stetson96}, and the Alarm variability statistic \citep{tamuz06}. Both Welch-Stetson $L$ and Alarm statistics are measures of coherent variability and make use of magnitude residuals relative to a weighted mean light curve magnitude. The implementations for calculating the Welch-Stetson $L$ and Alarm statistics were both provided by \textsc{Vartools}, an open source set of programs for analyzing time series data \citep{hartman16}. The Welch-Stetson $L$ statistic searches for variability by measuring the time-dependent correlation of residuals of magnitude pairs. The timespan for magnitude pairs was set to 10 days. The Alarm statistic was originally designed for detecting eclipsing binary star light curves and measures the correlation of adjacent residuals with the same magnitude sign. A large Alarm statistic value indicates that there are a large number of sequential residuals with the same sign which should correlate to coherent variability in the light curve. 

We have not applied a hard amplitude cut-off to our data set to separate Miras from SRs for multiple reasons. As discussed in \S \ref{sec:intro}, although SRs typically have smaller amplitudes than Miras, the conventional amplitude limit is arbitrary and high-amplitude SRs may have similar characteristics to Miras. Also, using an amplitude criterion with KELT data has problems. First, there exists no conventional amplitude cut-off for identifying Miras using the KELT bandpass. Second, yearly gaps in the observed data may result in incomplete coverage across a pulsation cycle of a star which hinders our ability to accurately measure amplitudes of some variables. Because of these difficulties, we find it hard to disentangle Miras from SRs in our catalog, and we note we have included in our catalog SRs with similar light curve characteristics to Miras. 

\subsubsection{Features from the Periodogram}\label{sec:rfperiodogram}

We also calculated the Lomb-Scargle (LS) periodogram \citep{zechmeister09,lomb76,scargle82} of the KELT light curves using \textsc{Astropy} \citep{astropy13, astropy18}. We tested the use of other period finding algorithms such as phase dispersion minimization \citep{stellingwerf78} and Analysis of Variance (AoV) period search \citep{schwarzenbergczerny89, devor05}, but found LS to be most effective. From the LS periodogram we used the following features for our RF classifier: the combined light curve LS period, the maximum normalized LS power, and a measure of the periodogram S/N.

We searched for periods between 50 days and half the baseline of the KELT object under analysis using a frequency spacing of $\Delta f = 10^{-5}$ days$^{-1}$ and used the maximum LS power to select periods. This period search was applied to the eastern, western, and combined light curves separately. We adopted the period from the combined data as the correct period, but verified the period was found in eastern and western data. We discuss cases in which the eastern and western periods disagreed requiring visual inspection later in \S \ref{sec:rfresults}. 

The LS power is a dimensionless value which measures how well a sinusoidal model of a particular frequency, amplitude, and phase fits the light curve data. It is normalized such that a sinusoidal model which perfectly fits the data would result in a power of one, and a constant non-varying model would result in a power of zero. The variables we are interested in have quasi-regular periodic signals; therefore, their periodograms should have higher LS powers compared to less-regular variable stars. There are distortions and power leakages in the power spectra of both Mira-like and non-Mira stars. Irregular cadence and gaps in the observations cause aliases in our data; variability departing from a perfect sinusoid can also cause harmonics of the true frequency to appear in the spectrum; and multiperiodicity, changing periods, and variable amplitudes can all increase the powers of spurious frequencies. We therefore determined the quality of the periodogram results by calculating a S/N for the max power similar to that used by \citet{benko07} and \citet{hartman16}. This is defined as $S/N_{LS} = (LS_{peak} - \langle LS \rangle)/{rms_{LS}}$, where $LS_{peak}$ is the maximum periodogram power, $\langle LS \rangle$ is the mean of the periodogram, and $rms_{LS}$ is the root mean square of the periodogram powers. This value compares the peak periodogram power to the background periodogram signal. Periodograms with low power leakage will have sharply defined peaks at the pulsation frequency (and at aliases and harmonics), and low power values at all other frequencies. This results in a high S/N value when compared to periodograms with higher power leakage which exhibit higher power values across multiple spurious frequencies. Mira-like variables, with their more regular periodic behavior, will exhibit periodograms with larger S/N values than irregular variables. We should note that the noise of a periodogram is not normally distributed; therefore, the S/N value does not follow a normal distribution. The S/N value is also sensitive to light curve baselines, as shorter baseline observations cause broader periodogram peaks which increases the value of $\langle LS \rangle$ and $rms_{LS}$. For each object we also calculated upper and lower error estimates on our periods using the FWHM of the periodogram peaks. We attempted to determine errors using Equ.\ 52 of \citet{vanderplas18}, but found this method underestimates errors, while the chosen method of using the periodogram peak FWHM overestimates errors.

\subsubsection{Training and Running Random Forest}\label{sec:rfrun}

After determining features for all objects passing the cuts discussed in \S \ref{sec:cuts}, we trained the RF classifier on a sample of VSX objects. We cross matched our KELT catalog to VSX stars classified as either Mira, SR, irregular, or miscellaneous (these are typically unclassified irregulars according to VSX) using a 6 arcsec search radius. We then retained only the VSX stars which passed the color, \textit{rms}, and amplitude cuts of \S \ref{sec:cuts}. We labeled the VSX Miras as Mira-like and inspected a sample of 299 high-amplitude ($\Delta R_{KELT} > 1.0$) VSX SR, irregular, and miscellaneous stars. Of these, we reclassified 172 as Mira-like as their amplitudes and near-regular periodicity were similar to lower-amplitude Mira light curves. We labeled the remaining SR, irregular, and miscellaneous objects as non-Miras. This resulted in a training set of 13,175 objects, 2453 (19\%) of which were classified as Mira-like.

We used \textsc{Scikit-Learn} to train a RF classifier of 500 decision trees, each with a maximum depth of 20, on the training set. The classifier returns probabilities of each object in the training set to be classified as Mira-like. To check for misclassified VSX objects in our training set we used a ten-fold cross-validation and found a classification score threshold of 0.5 resulted in a sample with 92\% purity and 88\% completeness. We adopted this score to create a preliminary sample of 255 false negatives (VSX Miras misclassified in the cross-validation set) and 182 false positives (VSX non-Miras misclassified as Miras in the cross-validation set). We visually inspected the light curves of these objects and reclassified 201 VSX objects. These new classifications were then introduced into our training set and the RF was re-trained. A ten-fold cross-validation was then used to determine that a classification score of 0.5 returned an improved sample of Mira-like objects with 94\% purity and 91\% completeness. We applied the trained RF classifier on KELT light curves passing the cuts described in \S \ref{sec:cuts}. Objects with classification scores greater than 0.5 were classified as Mira-like and selected for further study. The magnitudes for this classified sample were systematically fainter than the training set, as the training set mostly included well known, previously classified variables. Magnitudes do not have a direct affect on the RF classifier as they are a feature with low significance (see \S \ref{sec:rfphotometry}), but faint magnitudes can affect the values of other features. At fainter magnitudes light curves can exhibit more noise which can decrease variability statistics such as Welch-Stetson $L$ and Alarm. Fainter magnitude stars may also be more susceptible to blending from non-variables. This can decrease their light curve amplitudes. We are unable to correct for these in the KELT data, and note it as a limitation.

\section{Results}

\subsection{Results from Random Forest}\label{sec:rfresults}

Our RF classifier returned 4,491 Mira-like variables. Light curves of these Mira-like stars were selected for visual inspection if the selected frequency of the eastern, western, or combined periodograms was located at the edge of the frequency grid used (implying that the true period is outside the frequency grid); if either the eastern, western, or combined periods were greater than 1,000 days; or if the eastern, western, or combined frequencies differed by more than 0.0005 day$^{-1}$. These criteria resulted in an inspection of the light curves of 525 of the 4,491 Mira-like objects. Of these, 130 stars were removed from the catalog either because of ill-defined periods caused by irregularities in their light curve or poor time baseline coverage, 62 stars had their periods redetermined by selecting periodogram peaks which resulted in cleaner phase-folded plots, and 333 stars required no change in period as their combined light curve periods were deemed correct. 

We also attempted to identify and correct for issues due to blending by removing false positives due to contamination from nearby variables. Non-variable objects suffering from blending with a nearby variable have light curve periods near identical to that of the variable's period. We identified contaminated light curves by searching for groups of neighboring Mira-like objects in our catalog. Mira-like stars that were within 3 arcmin of each other and had a period difference less than 5\% were selected as a group. The object in each group with the brightest \textit{J} magnitude was kept and the remaining objects were removed from the catalog. This resulted in the removal of 229 objects from the Mira-like catalog. 

After visually inspecting the low quality set, correcting for blending, and removing the above mentioned stars, we produce a final catalog of 4,132 Mira-like objects. Henceforth, we refer to this catalog of 4,132 objects as the KELT Mira-like catalog. These are listed in Table 2. A sample of 14 light curves representative of our catalog are shown in Figure \ref{fig:sample_lcs}.

\subsection{Content and Completeness of Our Catalog}\label{sec:catalogs}

We compared our KELT Mira-like catalog to the catalog of Miras and SRs in VSX, as well as several other catalogs.  One of the difficulties in doing this is in identifying which stars in the other catalogs we would expect to be able to detect with KELT.  That is partly due to the different magnitude regimes probed by the different projects.  The VSX catalog lists the magnitude at maximum amplitude for each object, but these magnitudes are not consistently assembled from \textit{V}-band observations but from a variety of passbands. This hinders determining if a VSX object is within the KELT magnitude range. To obtain consistent \textit{V}-band magnitudes for the VSX sample, we matched the 15,073 VSX objects to the UCAC survey \citep{zacharias2013} using a 6 arcsec search radius and selected the nearest match. Not all VSX objects had UCAC \textit{V} magnitudes, but of the VSX objects which did we found approximately half (52\%) were within the KELT magnitude range (8 \textless{} \textit{V} \textless{} 13).

Of the 78,550 Miras and SRs in the VSX catalog with 2MASS-matched colors, 15,073 overlap with the KELT footprint, and we find 7,951 (52.8\%) of those in the KELT catalog.  This fraction is similar to the fraction of VSX objects in the KELT magnitude range using UCAC \textit{V} mangitudes, suggesting we are recovering most VSX objects within the KELT magnitude range.

For a further test of our cross-matching procedure, we inspected a sample of VSX objects which failed to match to KELT but had UCAC \textit{V} magnitudes within the KELT magnitude range. We found that nearly all failed matches were due to blending of the VSX variable with a nearby bright source in the KELT images. The remaining failed matches were due to the VSX object lying near the edge of a KELT field. The PSFs of these objects are distorted in KELT, causing errors in their astrometric positions in the KELT catalog.

We also found a small sample (180 objects) of dim VSX objects that match to a KELT object but have UCAC \textit{V} magnitudes outside the KELT magnitude range. These stars are faint in \textit{V} but are red enough to be detected by KELT's broad \textit{R} filter. It is also possible a portion may be incorrect matches between either VSX and KELT, or VSX and UCAC.

The 7,951 VSX Miras and SRs that match to stars in the KELT catalog can be broken up into 2,228 Miras and 5,723 SRs. Our KELT Mira-like catalog recovered 2,072 (93\%) of these VSX Miras, and only 623 (11\%) of these VSX SRS. This suggests our RF classifier performs well in identifying the Miras that exist in the KELT catalog. The RF classifier does not identify a large portion of the VSX SRs, though, but this should not be alarming as our study focuses on high-amplitude, regularly-periodic stars. Therefore, it is expected we exclude the majority of SRs which exhibit small-amplitude or irregular pulsations.

In addition to the VSX Miras and SRs, our Mira-like catalog also includes other classification types from VSX. We find 54 LPVs (long-period variables of unspecified type), 480 irregulars, 226 MISC stars (miscellaneous variable stars which failed to be classified by automatic analysis), and 59 VAR stars (variable stars of unspecified type) as identified by the VSX catalog. We also find 24 stars of various VSX classified types, with each type having fewer than 4 matches in our catalog, e.g. our catalog includes 3 RV Tauri stars, 3 UX Orionis stars, etc. The final matches between our KELT Mira-like catalog and VSX indicate that 58\% of our catalog are classified as Miras, 18\% as SRs, and 24\% as other variability types, but an unknown fraction of these assorted VSX types (e.g. LPV, MISC) could be reclassified as either Miras or SRs. 

For a better determination of the fraction of Miras and SRs in our KELT Mira-like catalog, we used the \citet{wozniak04b} catalog of red variables in NSVS (hereafter referred to as W04). The W04 catalog contains 8,678 red variable objects, 7,157 of which were within the KELT fields. We matched the W04 catalog to the KELT catalog using a 20 arcsec radius (the NSVS pixel size is 14.4 arcsec) and found KELT light curves for 5,437 objects. W04 classifies these objects into 416 carbon stars (C), 1,092 Miras (M), and 3,929 semi-regulars or irregulars (SR+L). Our KELT Mira-like catalog recovered 1,836 W04 objects: 179 are class C, 1,055 are class M, and 602 are class SR+L. Most W04 objects our KELT Mira-like catalog failed to recover are the low-amplitude SRs. We also found that most W04 class C objects showed irregular periodic behavior which is expected for carbon stars due to their asymmetric mass loss. This irregularity is the likely cause of the RF classifier not recovering the class C objects alongside the more regularly periodic Miras. Of the W04 objects, approximately 67\% are classified as class C or M, and 33\% are class SR+L. From this we assume that approximately 70\% of stars in our KELT Mira-like catalog are Miras, and 30\% are SRs.

The \citet{vogt16} catalog of Miras in ASAS (hereafter referred to as V16) contains 2,875 objects; 918 of which were within the KELT fields. These were matched to the KELT catalog using a 20 arcsec radius (the ASAS pixel size is 15 arcsec). We found light curves for 723 objects and successfully recovered 554 of the V16 Miras in our KELT Mira-like catalog. Every V16 Mira was found to have an VSX match. Those objects in either the W04 or V16 catalog with no KELT light curves are either outside the KELT magnitude limit, near the edges of the KELT fields, or blended with nearby objects.

We also compared our results to the \citet{oelkers18} catalog of 52,741 variable KELT objects. We have used an improved scheme of cross-matching KELT to 2MASS (see \S \ref{sec:crossmatch}) compared to \citet{oelkers18}, as they matched KELT to 2MASS by selecting the nearest positional match without incorporating color or magnitude information. We cross-matched our KELT Mira-like catalog to \citet{oelkers18} using a search radius of 50 arcsec. This radius was chosen to account for cases where the 2MASS identification, and therefore 2MASS coordinates, of KELT objects differed between our catalog and theirs.  Out of our 4,132 KELT Mira-like objects, 2,848 cross-matched to the \citet{oelkers18} catalog. Of these, 345 had periods listed by \citet{oelkers18}, and 30 of these periods differed from ours by more than 1\%. We found 1,255 KELT objects where the \citet{oelkers18} listed 2MASS identification differed from our catalog. 

We also matched our KELT Mira-like catalog to the \textit{Gaia} Data Release 2 (\textit{Gaia} DR2, \citealp{gaia18}) using a 1 arcsec search radius and selecting the nearest positional match, and found \textit{Gaia} DR2 matches for all 4,132 objects. We do not make use of the astrometric results from \textit{Gaia} DR2 in our study for several reasons, though. The huge size of Miras, which have variable diameters that are strongly wavelength dependent, result in angular diameters at least twice the size their parallaxes \citep{whitelock00}. Miras with thin dust shells have surface features that are dominated by convection and therefore have shifting photo-centers \citep{freytag17}. The astrometric errors introduced by this are discussed by \citet{chiavassa18} for SR variables and will be even larger for Miras. For Mira stars with appreciable dust shells, which are expected to be patchy, the uncertainties will be larger still \citep{whitelock12}. The large angular sizes combined with shifting photo-centers from asymmetric light distributions across the disk make the parallax measurements for AGBs currently unreliable. The large variation in magnitude and color of Miras also affects the accurate determination of parallaxes. Currently, \textit{Gaia} DR2 parallaxes are determined assuming constant magnitude and color for each source. This causes further inaccuracies for Mira parallaxes \citep{mowlavi18}. It may be possible to derive good parallaxes for Miras and \textit{Gaia}, but probably only at the end of the mission.

We also matched our KELT Mira-like catalog to the \textit{Gaia} DR2 Catalog of LPV Stars (\citealp{mowlavi18}; hereafter referred to as M18). The M18 catalog contains 151,761 LPV candidates with period measurements from \textit{Gaia} DR2, 23,405 of which M18 have classified into Miras. Using a 1 arcsec search radius and selecting the nearest positional match we matched our KELT Mira-like catalog to the M18 catalog and found matches for 2,440 objects. 

\begin{figure}[t]
\centering 
\subfloat{%
 \includegraphics[scale = .5]{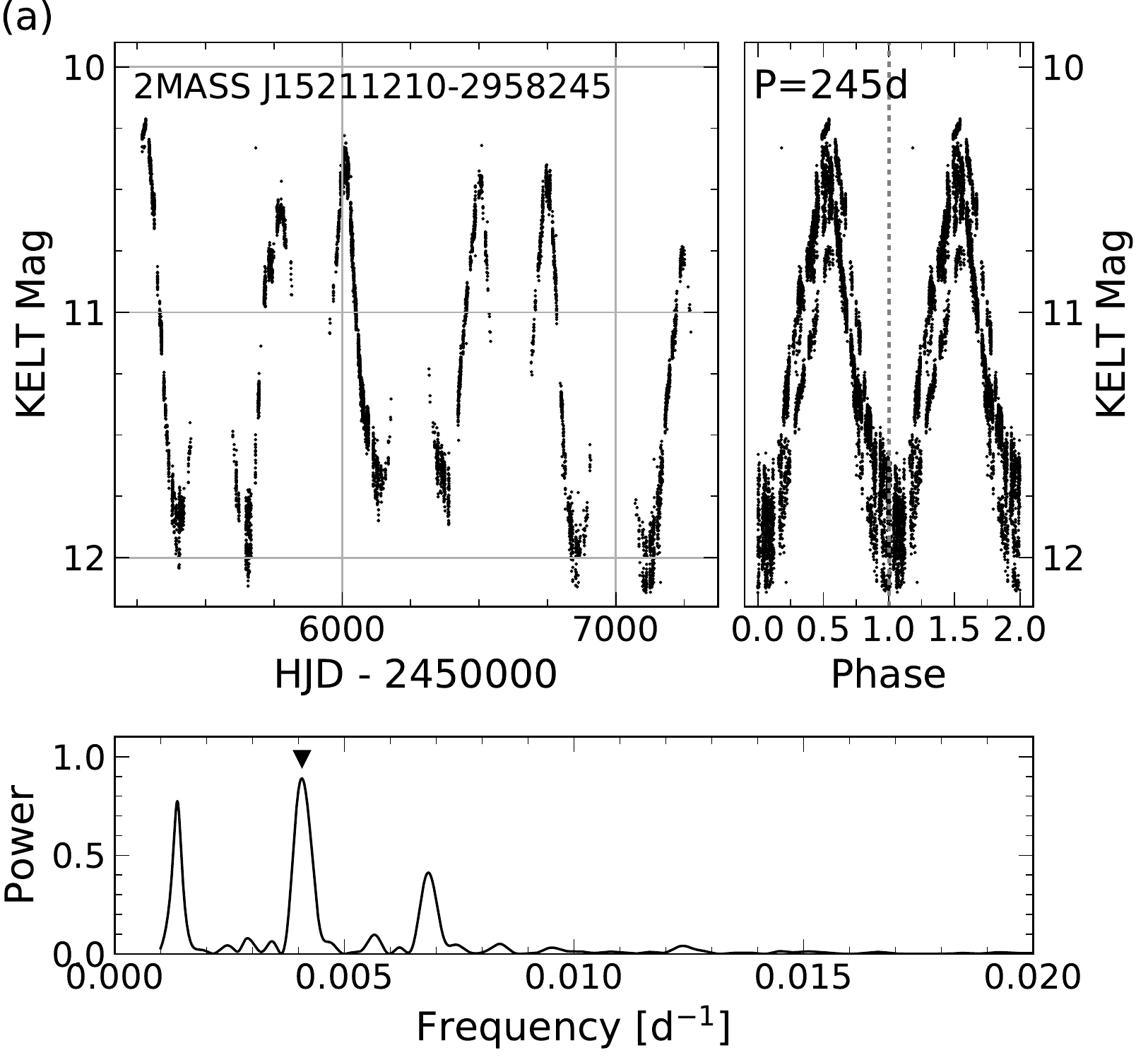}%
}
\qquad
\subfloat{%
 \includegraphics[scale = .5]{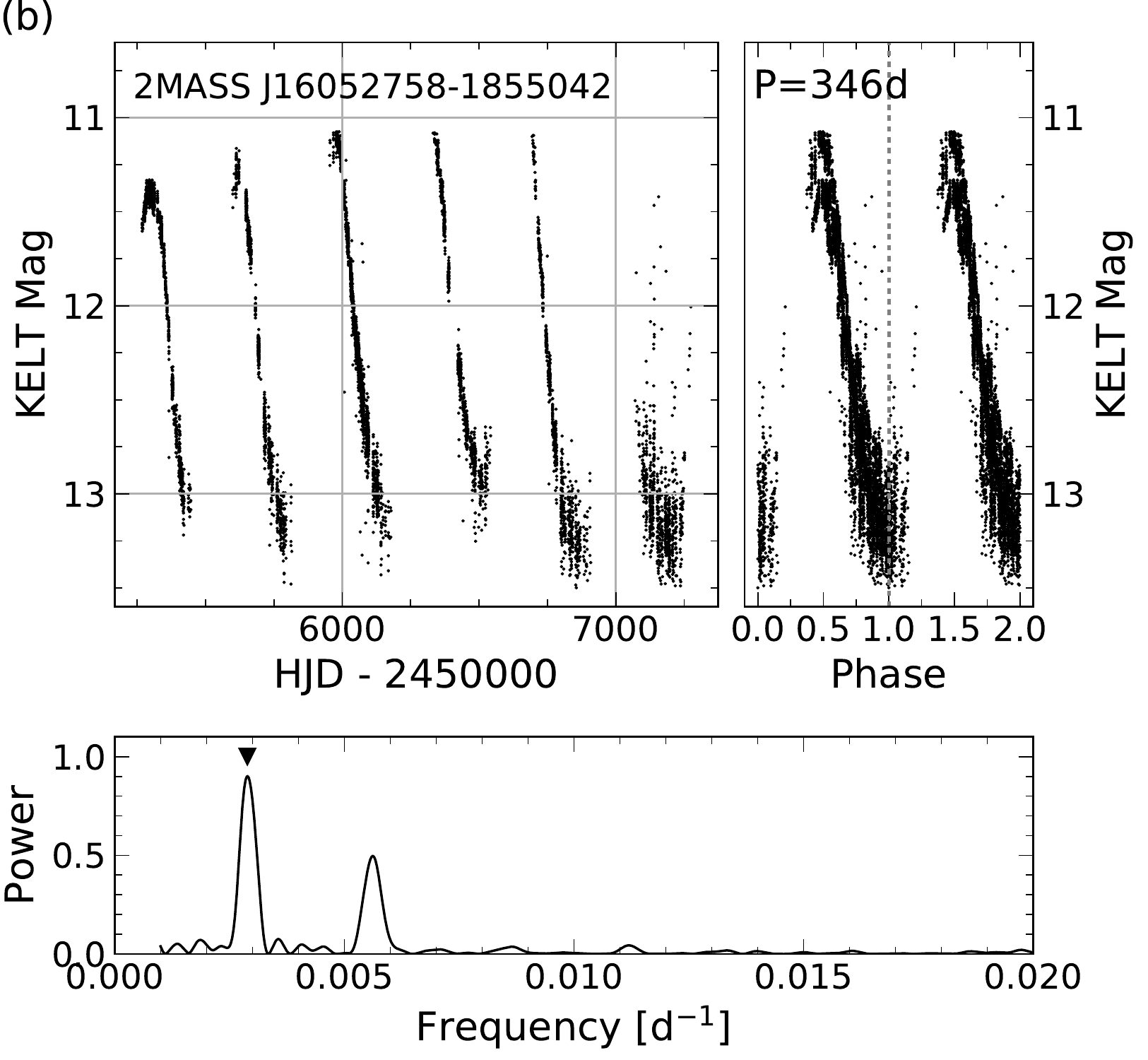}%
}
  \qquad
\subfloat{%
 \includegraphics[scale = .5]{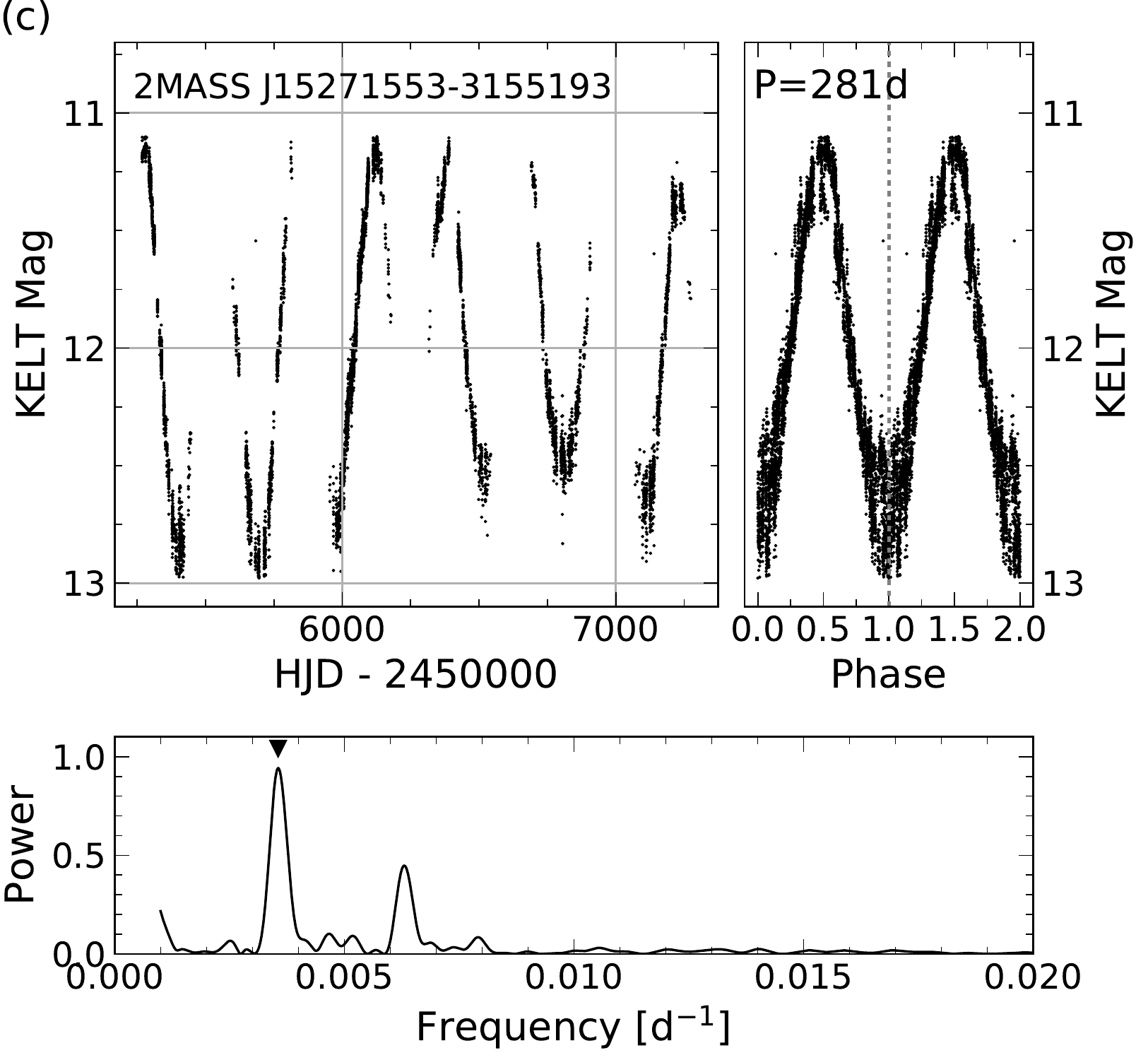}%
}
   \qquad
\subfloat{%
 \includegraphics[scale = .5]{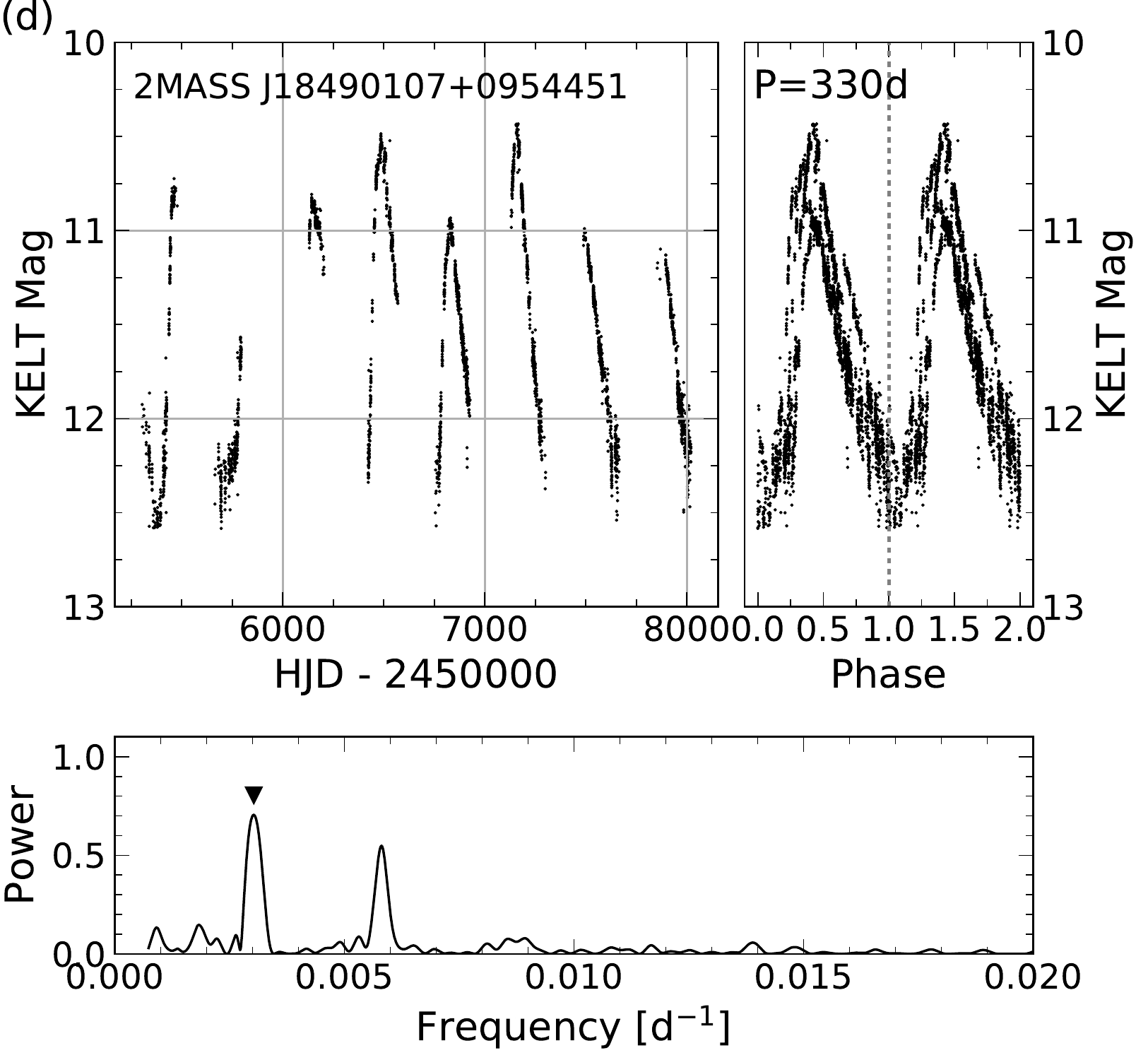}%
}
\caption{A sample of four KELT light curves from our catalog which have not been previously classified as variable. The 2MASS identifiers and measured periods are: (a) 2MASS J15211210-2958245 with P = 245d, (b) 2MASS J16052758-1855042 with P = 346d, (c) 2MASS J15271553-3155193 with P = 281d, (d) 2MASS J18490107+0954451 with P=330d.}
\label{fig:sample_new_lcs}
\end{figure}

We are particularly interested in knowing how many objects in our KELT Mira-like catalog are new Mira-like detections. In addition to VSX, we made use of the W04 catalog and the SIMBAD database. We do note that SIMBAD is incomplete, and our inclusion of the VSX and W04 catalog attempts to mitigate this. Our KELT Mira-like catalog contains 996 objects which failed to match to a known Mira or SR, from either VSX or W04. These failed matches were matched to SIMBAD using a 2 arcsec search radius. For cases in which a coordinate search to SIMBAD produced no results, we attempted a SIMBAD match using the 2MASS ID. We used this sample to determine the classification status of our Mira-like stars. Of the objects in our KELT Mira-like catalog not classified as Mira or SR by VSX or as a red variable by W04, the SIMBAD classifications state 116 are AGB (or related type) stars; 66 are candidate or known Miras, LPVs, or SRs; and 814 matches had either unrelated SIMBAD classifications or no match. Figure \ref{fig:sample_new_lcs} shows light curves of four stars not previously classified as variable.

 \begin{figure}[t]
\centering 
\subfloat{%
 \includegraphics[scale = .4]{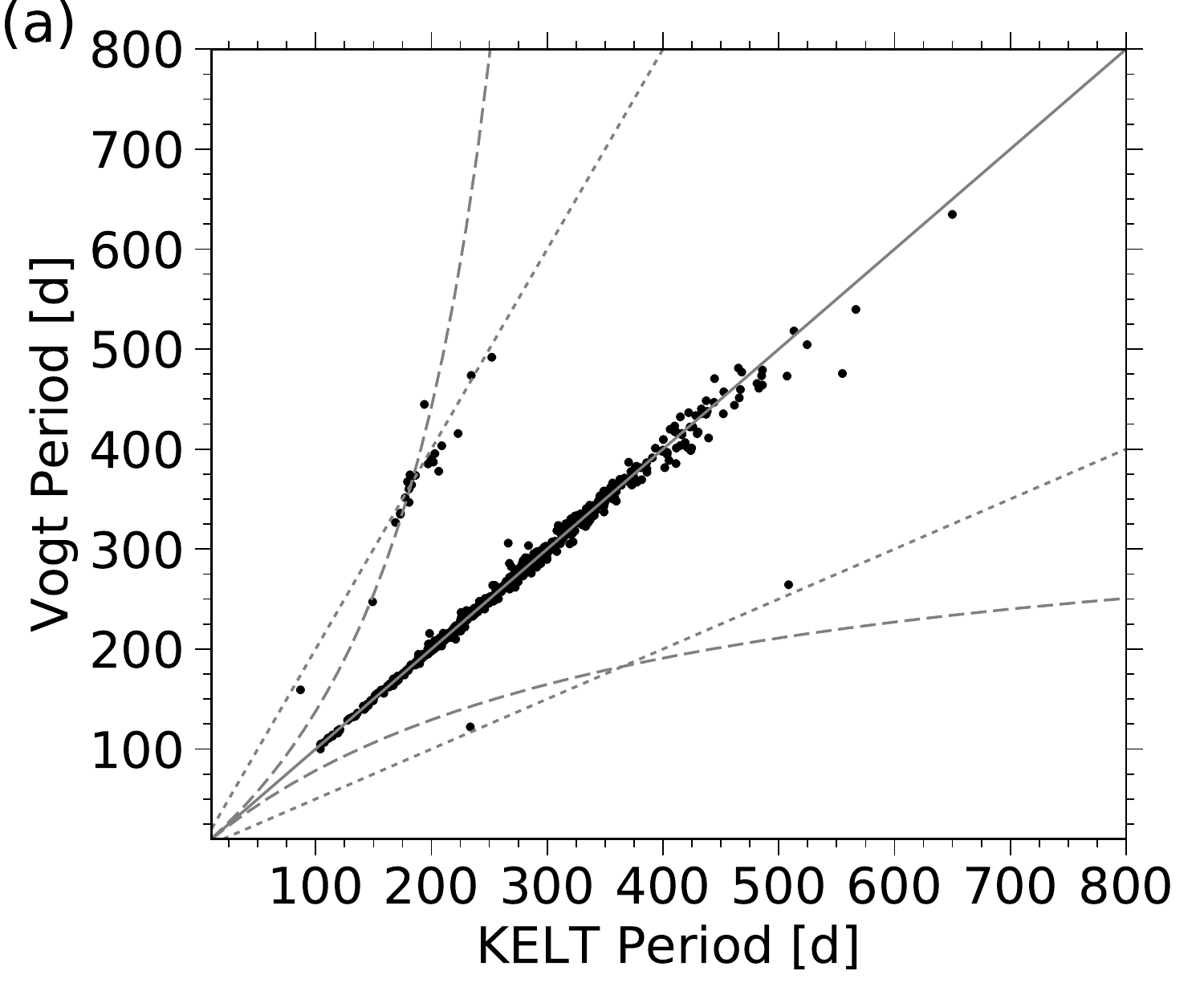}
}
\subfloat{%
 \includegraphics[scale = .4]{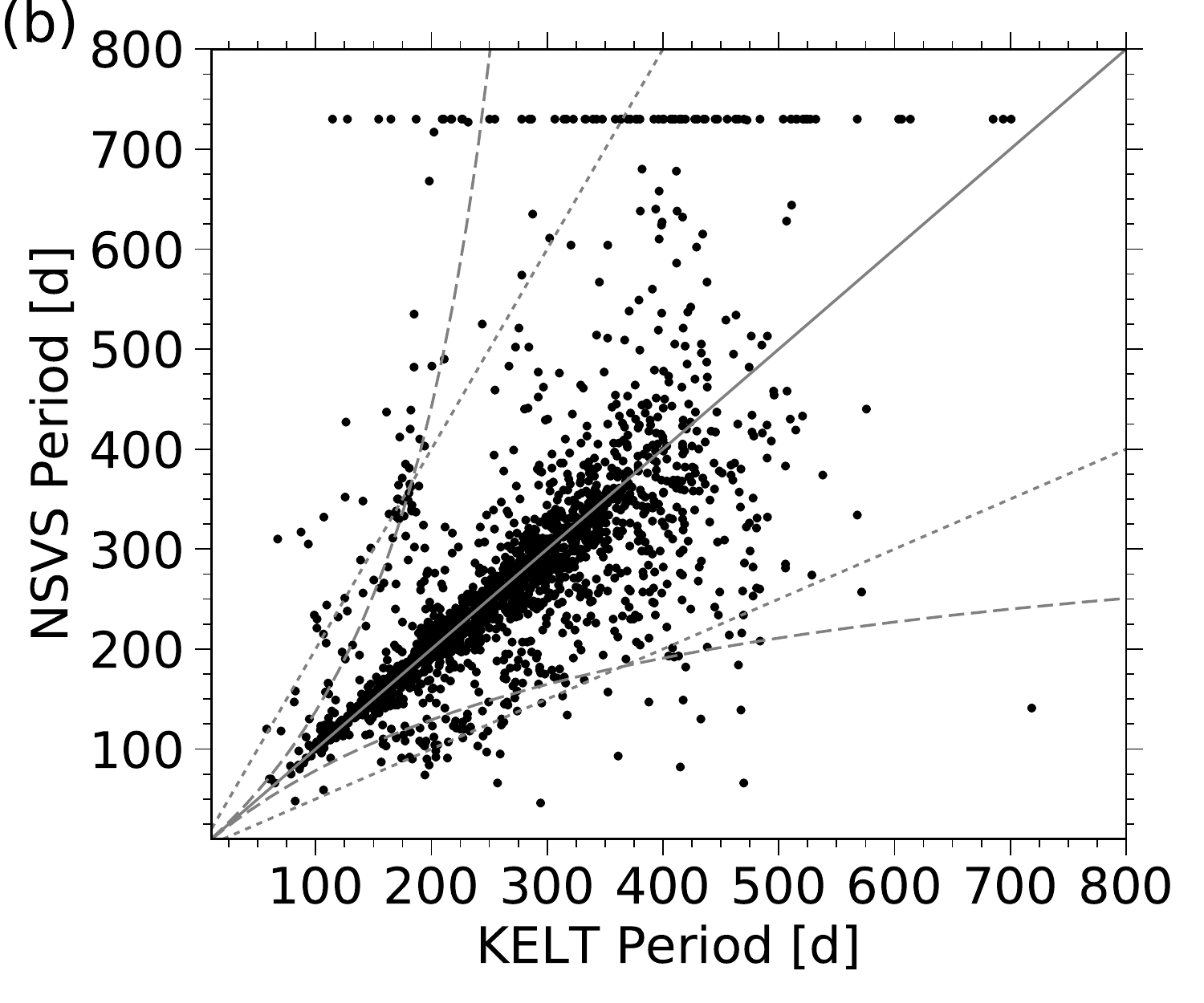}
}
\subfloat{%
 \includegraphics[scale = .4]{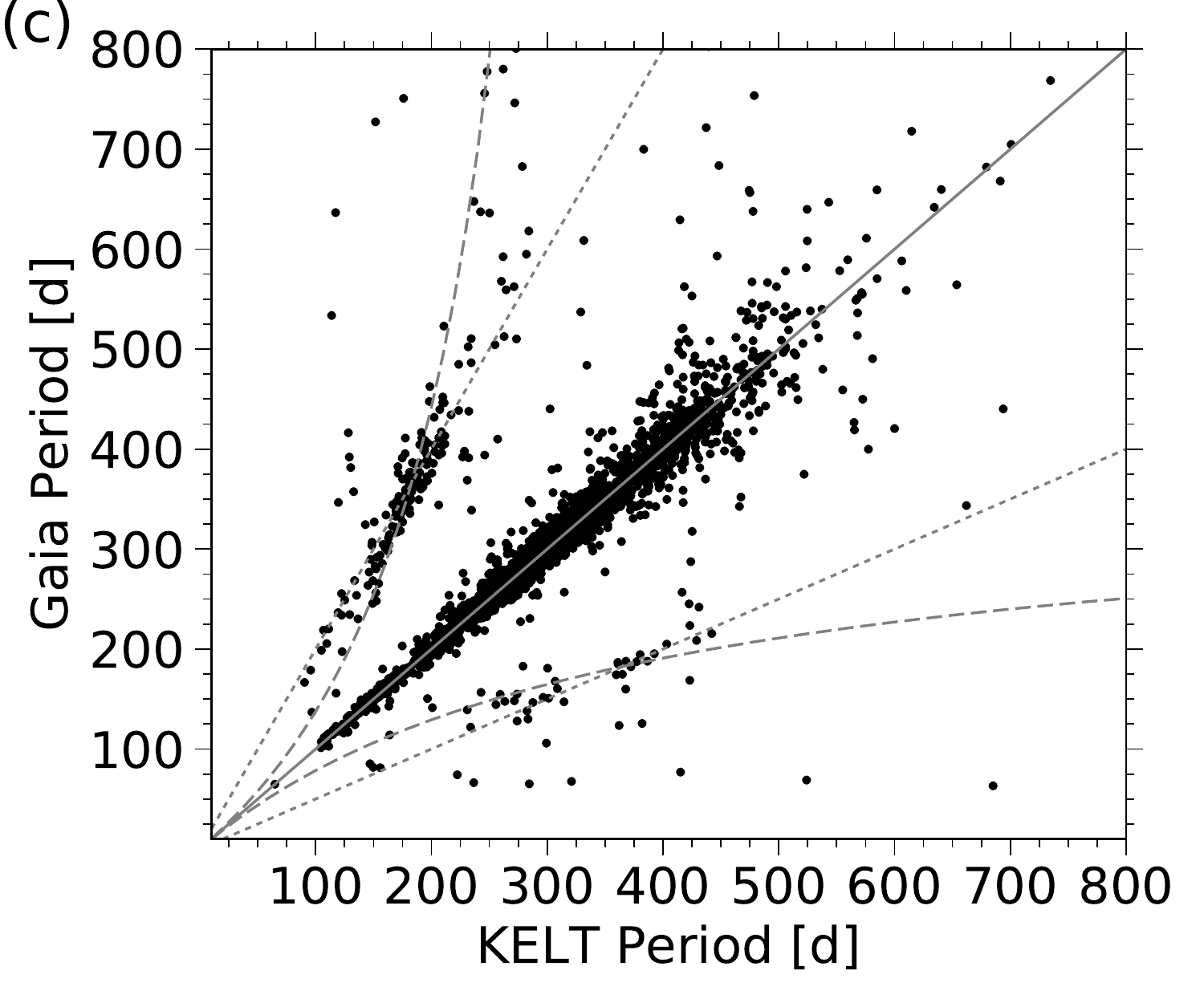}
}
\caption{We matched our Mira-like catalog to the catalogs of (a) V16, (b) W04, and (c) M18 to verify the accuracy of our measured periods. Predicted aliases are shown as dashed lines, and period harmonics are shown as straight dotted lines. We find good agreement between our periods and those of V16 with most discrepancies near the harmonics. The periods of W04 appear to be less reliable likely due to the shorter time baseline of their work. There is also good agreement between our periods and those of M18, but the number of discrepancies along aliases and harmonics is larger.}
\label{fig:period_comp}
\end{figure}

\begin{sidewaystable}[h]
\caption{Main properties of the first 20 stars in our catalog. The complete table is available online.}
\centering
\resizebox{\textwidth}{!}{
	\begin{tabular}{c c c c c c c c c c c c c c}
	\hline\hline
	 Eastern KELT I.D. & Western KELT I.D. & 2MASS I.D. & RA & Dec. & $l$ & $b$ & $R_{KELT}$ & $J$ & $H$ & $K_s$ & Period & $S/N_{LS}$ & mismatch flag \\ [0.1ex]
	 & & & (J2000) & (J2000) & ($^{\circ}$) & ($^{\circ}$) & (mag) & (mag) & (mag) 	& (mag)   & (days) & &  \\  
	 \hline
	
	KELT N01 lc 012200 V01 east & KELT N01 lc 039679 V01 west & 2MASS J23573235+3202204 & 23:57:32.36 & +32:02:20.48 & 109.820255 & -29.456737 & 11.05 & 06.08 & 05.30 & 04.88 & $227^{+7}_{-7}$ & 06.19 & 0 \\
	KELT N01 lc 012821 V01 east & KELT N01 lc 046147 V01 west & 2MASS J00594645+2756445 & 00:59:46.46 & +27:56:44.58 & 125.176497 & -34.889698 & 10.19 & 07.18 & 06.38 & 05.95 & $187^{+8}_{-4}$ & 06.05 & 0 \\
	KELT N01 lc 029631 V01 east & KELT N01 lc 048055 V01 west & 2MASS J23535704+3310562 & 23:53:57.05 & +33:10:56.27 & 109.302805 & -28.163464 & 12.18 & 08.18 & 07.25 & 06.92 & $581^{+72}_{-47}$ & 05.26 & 0 \\
	KELT N01 lc 033544 V01 east & KELT N01 lc 009353 V01 west & 2MASS J23514408+2945435 & 23:51:44.09 & +29:45:43.56 & 107.755008 & -31.344054 & 09.07 & 04.62 & 03.75 & 03.14 & $322^{+29}_{-17}$ & 05.70 & 0 \\
	KELT N01 lc 033596 V01 east & KELT N01 lc 023249 V01 west & 2MASS J00234610+3830077 & 00:23:46.11 & +38:30:07.73 & 117.007674 & -24.051108 & 10.88 & 08.92 & 08.13 & 07.93 & $412^{+35}_{-28}$ & 06.04 & 1 \\
	KELT N01 lc 041955 V01 east & KELT N01 lc 030143 V01 west & 2MASS J00500630+3539098 & 00:50:06.31 & +35:39:09.89 & 122.627460 & -27.218213 & 09.36 & 05.48 & 04.69 & 04.27 & $260^{+11}_{-10}$ & 05.94 & 0 \\
	KELT N01 lc 043787 V01 east & KELT N01 lc 046756 V01 west & 2MASS J00141091+2901203 & 00:14:10.92 & +29:01:20.33 & 113.200899 & -33.144494 & 10.39 & 05.69 & 04.84 & 04.44 & $238^{+8}_{-11}$ & 05.77 & 0 \\
	KELT N01 lc 046801 V01 east & nan & 2MASS J00471892+3241083 & 00:47:18.93 & +32:41:08.38 & 121.928508 & -30.178056 & 10.66 & 03.05 & 02.23 & 01.76 & $429^{+30}_{-34}$ & 05.58 & 0 \\
	KELT N01 lc 047788 V01 east & nan & 2MASS J00512328+3422363 & 00:51:23.29 & +34:22:36.32 & 122.920217 & -28.494993 & 10.89 & 04.88 & 03.84 & 03.25 & $329^{+21}_{-18}$ & 05.16 & 0 \\
	KELT N01 lc 051102 V01 east & nan  & 2MASS J00030086+3038233 & 00:03:00.86 & +30:38:23.34 & 110.768735 & -31.085668 & 13.41 & 10.90 & 09.71 & 08.90 & $306^{+16}_{-13}$ & 05.74 & 0 \\
	KELT N01 lc 051823 V01 east & KELT N01 lc 045548 V01 west & 2MASS J23491137+2638088 & 23:49:11.37 & +26:38:08.89 & 106.078971 & -34.190596 & 09.22 & 05.33 & 04.20 & 03.80 & $230^{+8}_{-6}$ & 06.37 & 0 \\
	KELT N01 lc 052269 V01 east & KELT N01 lc 017514 V01 west & 2MASS J23461773+3024468 & 23:46:17.74 & +30:24:46.88 & 106.650150 & -30.390832 & 11.08 & 06.04 & 05.14 & 04.75 & $129^{+3}_{-2}$ & 06.30 & 0 \\
	KELT N01 lc 053684 V01 east & KELT N01 lc 058300 V01 west & 2MASS J00322275+2601459 & 00:32:22.76 & +26:01:45.94 & 117.594322 & -36.644720 & 07.80 & 03.42 & 02.54 & 02.13 & $310^{+15}_{-14}$ & 07.05 & 0 \\
	KELT N01 lc 053981 V01 east & KELT N01 lc 048462 V01 west & 2MASS J00222314+2659458 & 00:22:23.15 & +26:59:45.83 & 114.985647 & -35.427023 & 08.69 & 04.68 & 03.92 & 03.50 & $279^{+14}_{-13}$ & 05.86 & 0 \\
	KELT N01 lc 054306 V01 east & KELT N01 lc 049456 V01 west & 2MASS J00482082+2703259 & 00:48:20.83 & +27:03:25.98 & 122.083382 & -35.809439 & 10.99 & 05.13 & 04.17 & 03.81 & $344^{+15}_{-32}$ & 05.18 & 0 \\
	KELT N01 lc 056383 V01 east & KELT N01 lc 045232 V01 west & 2MASS J23430657+3528452 & 23:43:06.58 & +35:28:45.21 & 107.583772 & -25.347799 & 10.89 & 04.92 & 03.91 & 03.24 & $377^{+31}_{-11}$ & 07.09 & 0 \\
	KELT N01 lc 058729 V01 east & KELT N01 lc 058434 V01 west & 2MASS J00000657+2553112 & 00:00:06.57 & +25:53:11.23 & 108.712957 & -35.564066 & 08.01 & 02.23 & 01.32 & 00.92 & $321^{+27}_{-14}$ & 06.18 & 0 \\
	KELT N01 lc 060087 V01 east & KELT N01 lc 058910 V01 west & 2MASS J23505191+4304452 & 23:50:51.92 & +43:04:45.21 & 111.329781 & -18.423770 & 12.71 & 07.67 & 06.84 & 06.50 & $280^{+11}_{-12}$ & 06.33 & 0 \\
	KELT N01 lc 065529 V01 east & KELT N01 lc 059071 V01 west & 2MASS J00042008+4006356 & 00:04:20.08 & +40:06:35.64 & 113.249082 & -21.874575 & 08.67 & 03.57 & 02.59 & 02.08 & $315^{+13}_{-16}$ & 07.11 & 0 \\
	KELT N01 lc 066205 V01 east & KELT N01 lc 049416 V01 west & 2MASS J23572739+2407312 & 23:57:27.39 & +24:07:31.21 & 107.440420 & -37.117208 & 11.53 & 06.59 & 05.72 & 05.25 & $244^{+10}_{-6}$ & 06.78 & 0 \\
	 \hline
	 \end{tabular}
 }
 \end{sidewaystable}
 
 \subsection{Comparing Periods to Other Catalogs}\label{sec:periods}

We compared the periods of variability we found to those of V16, W04, and M18. V16 used a non-parametric algorithm to re-analyze periods of Miras in the ASAS catalog, which required human interaction at certain stages. The KELT and V16 periods are compared in Figure \ref{fig:period_comp}(a) and agree very well. The W04 catalog has a short time baseline nearing 1 year for the longest observations. Therefore, they were unable to obtain precise periods for many of their Miras. There is a sample of Miras with periods given by W04 as 730 days (their maximum period range) which we have redetermined. Figure \ref{fig:period_comp}(b) compares the KELT and W04 periods. Figure \ref{fig:period_comp}(c) compares the KELT Mira-like and M18 periods. We see most periods agree well, but there is more scatter, and approximately 160 stars lie along the common aliases and harmonics. Many of these stars exhibit periods near a year or half a year which can demonstrate spurious signals due to systematic noise on annual timescales in the KELT data. 

\begin{figure}[t]
\centering 
\includegraphics[scale = .5]{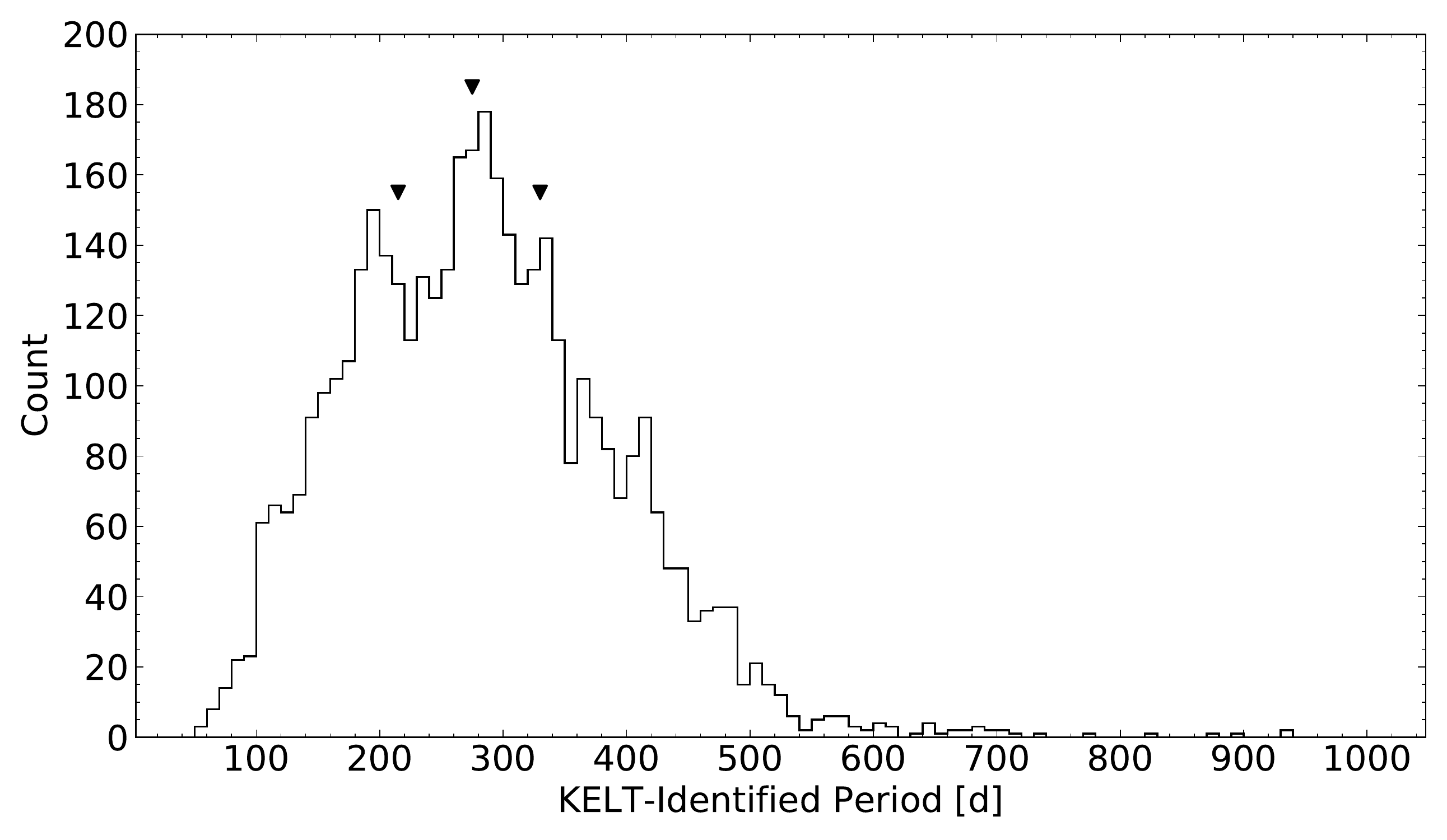}
\caption{Distribution of periods for our Mira catalog. Period values associated with peaks identified by V16 at 215, 275, and 330 days are marked by triangles. We find that the our peak at 215 days is contaminated by mismatched periods, and the 330 day peak is consistent with noise. We therefore cannot confirm the significance of the 215 and 330 day peaks.}
\label{fig:period_dist}
\end{figure}

V16 identified three peaks in their period distribution at 215, 275, and 330 days. When binning the data into 10 day bins, as done in V16, we find similar peaks in our distribution at 195, 285, and 335 days. Our period distribution is shown in Figure \ref{fig:period_dist} and the periods associated with the V16 identified peaks are represented by black markers. However, we find that the combination of systematic noise and aliasing of periods is responsible for the 195 and 335 day peak, and only the central peak at 285 days is significant.

\begin{figure}[t]
\centering
\includegraphics[scale = .5]{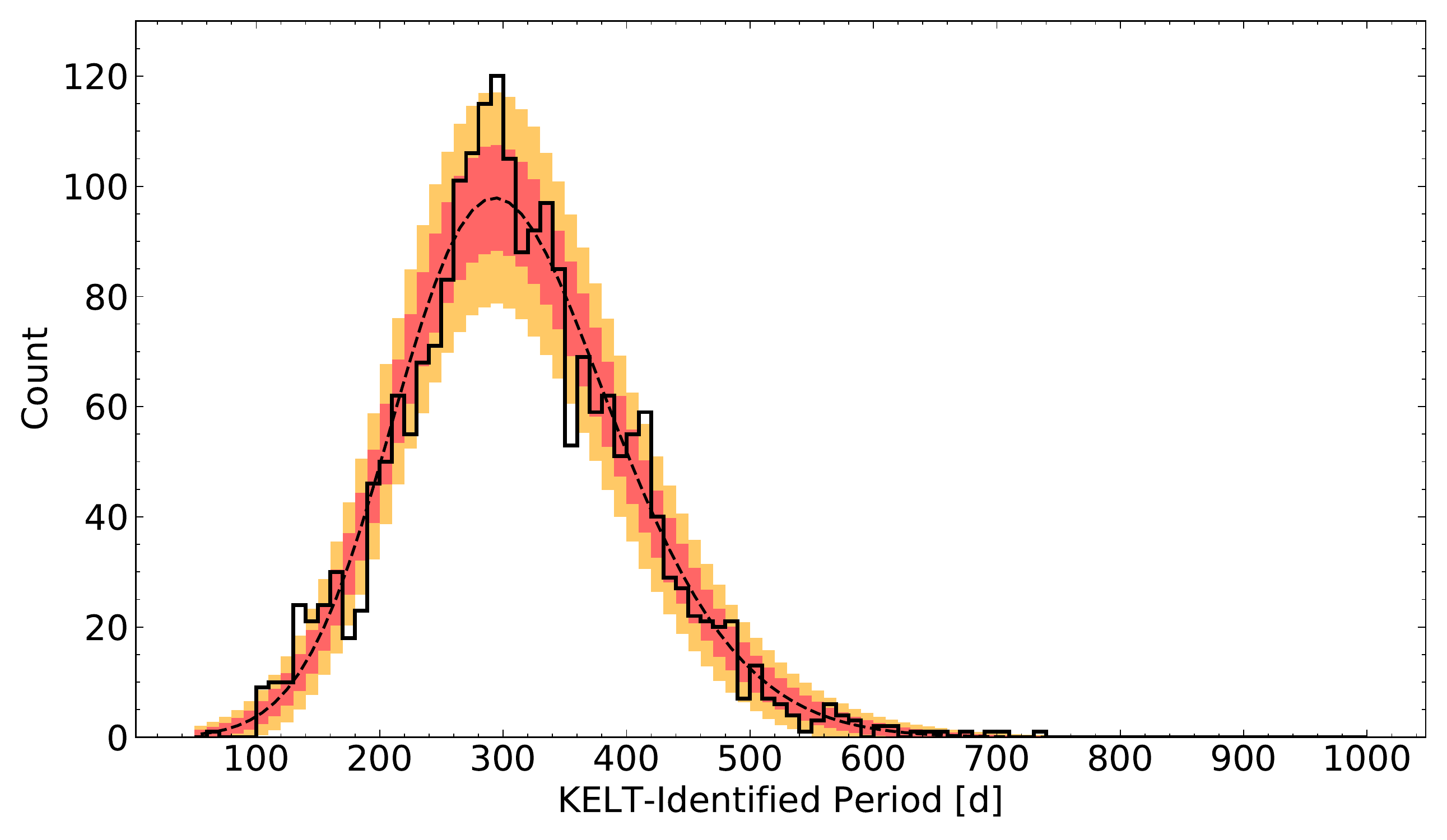}
\caption{Distribution of periods for stars in which our periods and M18 periods agree within 30\%. The dashed line represents a skewed Gaussian distribution function which has been fit to the data. The 1-sigma and 2-sigma confidence levels are shown represented by the red and orange areas respectively. We see that the 195 day peak disappears because of the removal of mismatched periods, and the peak near 335 day peak is consistent with noise.}
\label{fig:bin_error}
\end{figure}

To further analyse these peaks, we created a new histogram using only the stars in which our periods agreed with those of M18 within 30\%. This distribution is shown in Figure \ref{fig:bin_error}. We see that the 195 day peak disappears because of the removal of mismatched periods. We fit both a Gaussian and skewed Gaussian distribution function to this histogram and found negligible differences in the fit quality. The dashed line in Figure \ref{fig:bin_error} represents a skewed Gaussian fit to the histogram, and the 1-sigma and 2-sigma errors are represented by red and orange shading. These errors were calculated using the square root of the number of bin counts expected from the distribution function. It is apparent that the 335 day peak in this sample does not appear to be significant. Although a Gaussian is not a perfect fit, the distribution of the entire KELT Mira-like catalog is consistent with a Gaussian profile with mean period of 285 days and standard deviation of 111 days, and the additional peaks at 215 and 330 days claimed by V16 are not confirmed.

\begin{figure}[t]
\centering 
\includegraphics[scale = .5]{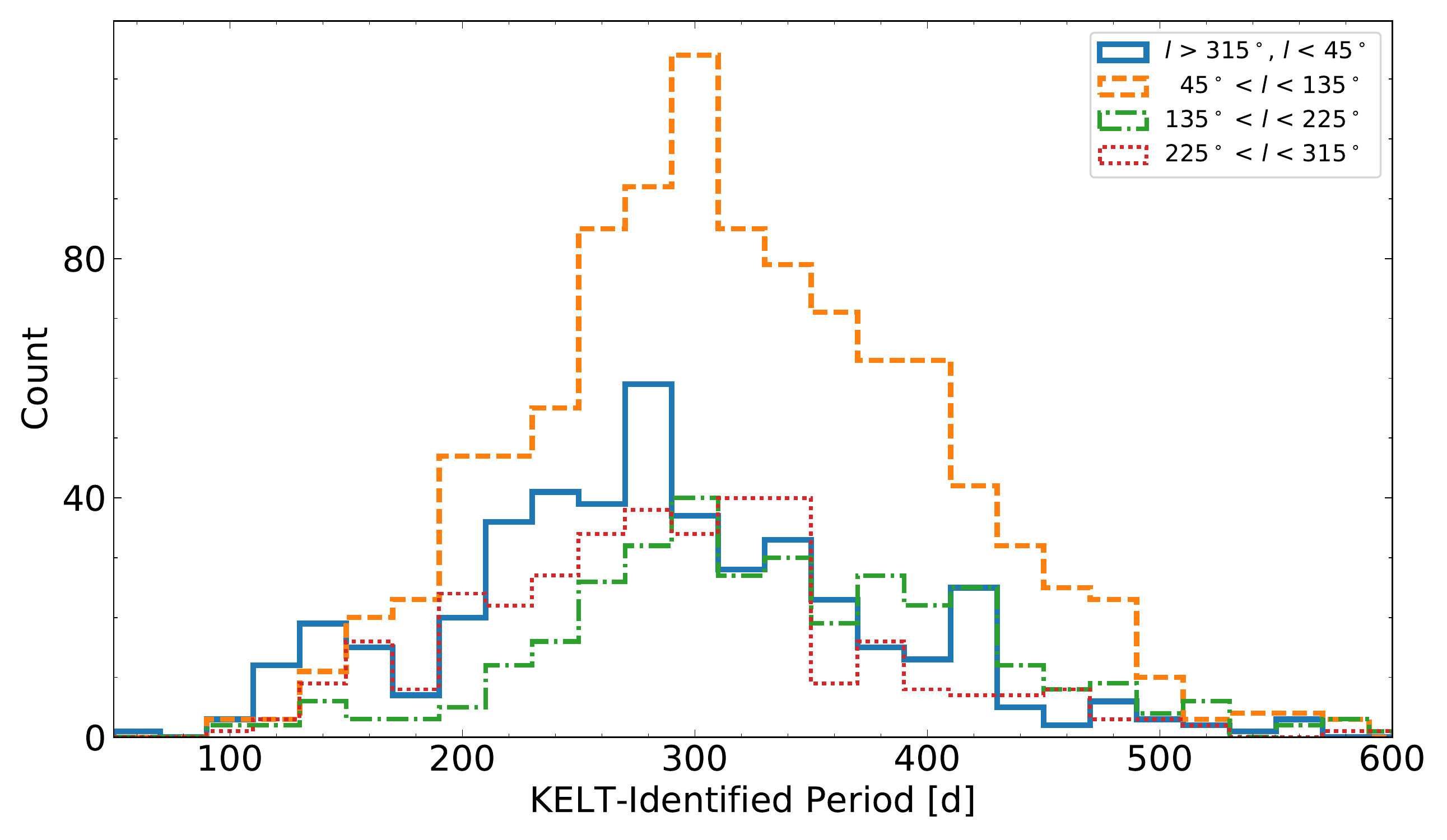}
\caption{Distribution of periods for stars in which periods agreeing within 30\% of M18 for four Galactic longitude bins. The solid blue histogram is for Miras between \textit{l} \textgreater{} $315^{\circ}$ and \textit{l} \textless{} $45^{\circ}$. The dashed orange histogram is for Miras between $45^{\circ}$ \textless{} \textit{l} \textless{} $135^{\circ}$. The dotted-dashed green histogram is for Miras between $135^{\circ}$ \textless{} \textit{l} \textless{} $225^{\circ}$, and the dotted red histogram is for Miras between $225^{\circ}$ \textless{} \textit{l} \textless{} $315^{\circ}$.}
\label{fig:period_dist_gal_long}
\end{figure}

We investigated the dependence of our period distribution on Galactic longitude and latitude. We have used only the sample of stars with KELT periods agreeing within 30\% of the M18 catalog to remove the peaks at 215 and 330 days. Figure \ref{fig:period_dist_gal_long} shows histograms of the period distributions for four longitude bins. Here we have increased the size of the period bins to 20 days to reduce counting errors. The solid blue histogram shows the periods of stars that lie in the range \textit{l} \textgreater{} $315^{\circ}$ and \textit{l} \textless{} $45^{\circ}$ (toward the Galactic center). The dashed orange histogram shows stars between $45^{\circ}$ \textless{} \textit{l} \textless{} $135^{\circ}$. The dotted-dashed green histogram shows stars between $135^{\circ}$ \textless{} \textit{l} \textless{} $225^{\circ}$ (toward the Galactic anti-center), and the dotted red histogram shows stars between $225^{\circ}$ \textless{} \textit{l} \textless{} $315^{\circ}$. V16 suggests, using arguments from \citet{feast14}, that the period distribution reflects the star formation history of the sampled regions, with shorter periods associated with older stars. Previous studies have shown that oxygen-rich (O-rich) Miras are concentrated toward the Galactic center, and that carbon-rich (C-rich) have a more uniform distribution across the Galaxy (\citealp{blanco65, noguchi04, ishihara11} and references therein) and O-rich Miras typically have shorter periods than C-rich Miras. However, we do not confirm a relation between Galactic longitude and KELT periods.

\begin{figure}[t]
\centering 
\includegraphics[scale = .5]{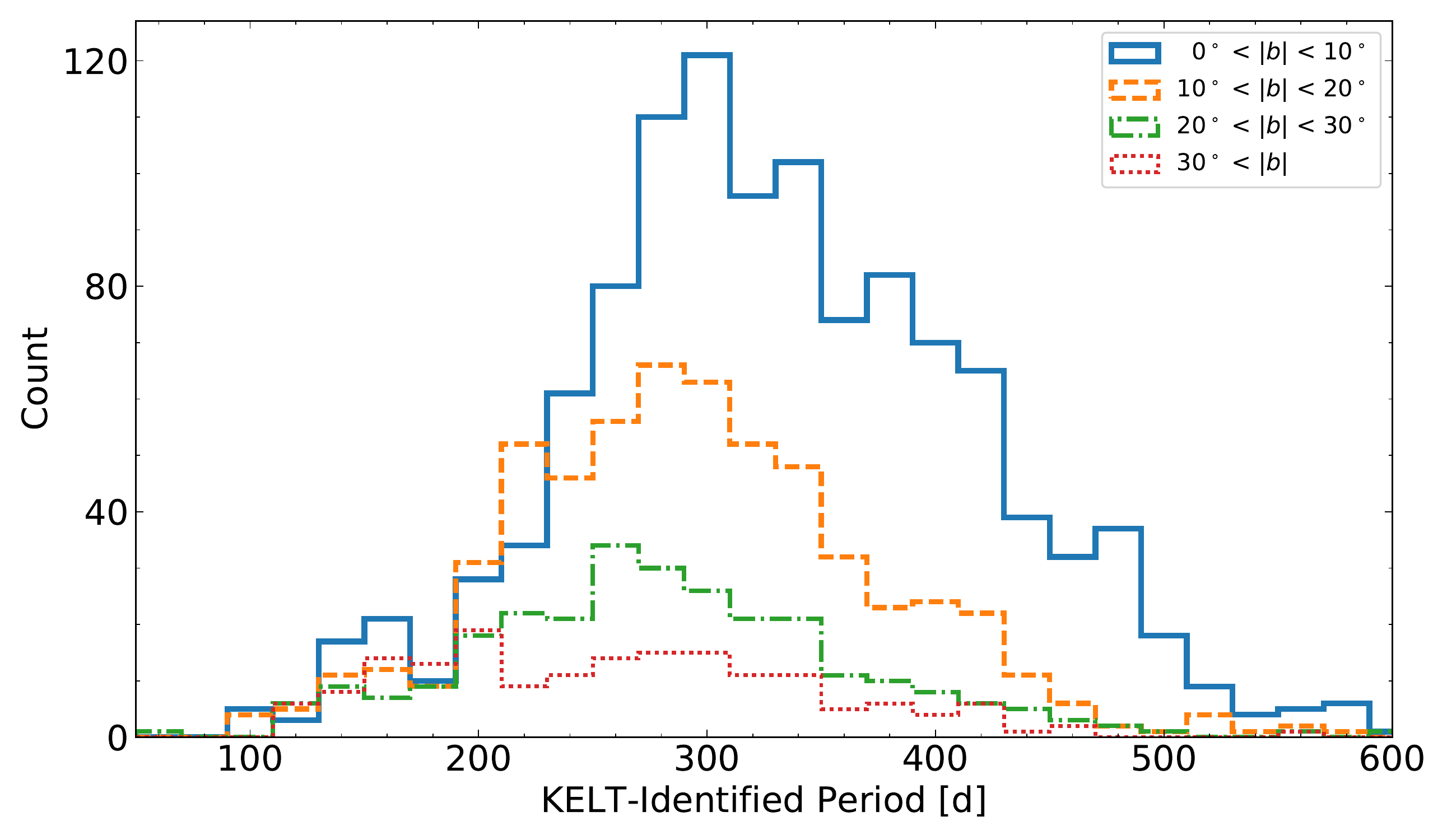}
\caption{Distribution of periods for stars in which periods agreeing within 30\% of M18 for four absolute Galactic latitude bins. The solid blue histogram is for Miras with $\lvert b \rvert$ \textless{} $10^{\circ}$. The dashed orange histogram is for Miras between $10^{\circ}$ \textless{} $\lvert b \rvert$ \textless{} $20^{\circ}$. The dotted-dashed green histogram is for Miras between $20^{\circ}$ \textless{} $\lvert b \rvert$ \textless{} $30^{\circ}$, and the dotted red histogram is for Miras with $\lvert b \rvert$ \textgreater{} $30^{\circ}$. We observe a shift toward lower periods at high latitudes.}
\label{fig:period_dist_gal_lat}
\end{figure}

Figure \ref{fig:period_dist_gal_lat} shows period histograms for four latitude bins, still using only KELT periods agreeing within 30\% of M18. The solid blue histogram is for Miras with $\lvert b \rvert$ \textless{} $10^{\circ}$. The dashed orange histogram is for Miras between $10^{\circ}$ \textless{} $\lvert b \rvert$ \textless{} $20^{\circ}$. The dotted-dashed green histogram is for Miras between $20^{\circ}$ \textless{} $\lvert b \rvert$ \textless{} $30^{\circ}$, and the dotted red histogram is for Miras with $\lvert b \rvert$ \textgreater{} $30^{\circ}$. Generally we find a shift toward shorter periods at higher latitudes. This is consistent with the expectation that longer period stars have higher mass progenitors and are therefore more concentrated in the Galactic plane. A similar effect can be seen in Figs.\ 16 and 17 from \citet{whitelock94}.
 
\section{Determining Carbon- and Oxygen-Rich Types}\label{sec:OvsC}

Mira variables fall into three subclasses depending on the abundances of oxygen vs.\ carbon in their atmospheres. Miras with similar oxygen and carbon abundances are referred to as S-type, and Miras with an imbalance between oxygen and carbon are referred to as either oxygen-rich (O-rich) or carbon-rich (C-rich). C-rich AGBs are typically redder than O-rich ones due to the reddening by dust in their atmospheres, thus they typically have redder $J - K_s$ colors \citep{feast82,marigo03}. Distinguishing O-rich from C-rich Miras is useful for distance studies. Compared to O-rich Miras, the PL relation in $K_s$-band breaks down over a longer range of periods for C-rich Miras due to circumstellar reddening caused by their thicker shells \citep{whitelock12}. This results in O-rich Miras having a better defined PL relation.

$J-K_s$ colors have commonly been used to distinguish O- from C-rich Miras. \citet{cole02} used the criterion of ($J-K_s$) \textgreater{} 2 to classify C-rich stars in the Bulge. However, these results have been questioned, and resulting studies suggest that $J-K_s$ colors alone are not accurate in distinguishing between C-rich and O-rich Miras \citep{ojha07, uttenthaler15, catchpole16}. Alternative diagnostics in selecting O-rich and C-rich from photometry have been explored. \citet{ishihara11} and \citet{matsunaga17} have used color-color diagrams of $J - K_s$ and \textit{AKARI} colors to select C-rich stars in the Bulge, and have shown this to be an improvement over the use of single 2MASS color. \citet{lebzelter18} have had success distinguishing subclasses of AGBs in the Magellanic Clouds using colors from both 2MASS and the \textit{Gaia} DR2. They were able to distinguish C-rich from O-rich Miras by plotting $K_s$ against a combination of 2MASS and \textit{Gaia} colors known as Wesenheit indices. The Wesenheit index $W_{K_s}$ is a reddening-free combination of 2MASS magnitudes defined as
\begin{equation}
W_{K_s} = K_s - 0.67 \ (J - K_s).
\end{equation}
This is derived assuming an $R_V = 3.1$ \citep{mathis90, dutra02, soszynski13}. The index $W_{RP}$ is a reddening-free combination of \textit{Gaia} magnitudes derived by \citet{lebzelter18},
\begin{equation}
W_{RP} = G_{RP} - 1.3 \  (G_{BP} - G_{RP}),
\end{equation}
where $G_{BP}$ and $G_{RP}$ are \textit{Gaia} blue and red apparent magnitudes respectively. Because the Miras in our catalog are Galactic and their distances uncertain, we cannot replicate their method of using $K_s$ in place of absolute magnitudes. Instead, we investigate the combination of Wesenheit indices ($W_{RP} - W_{K_s}$) and ($J - K_s$) colors. The color index ($W_{RP} - W_{K_s}$) is not a color in the classical sense, but a combination of three colors. In this index, bluer objects are found near ($W_{RP} - W_{K_s}$) = 0.8 and redder objects at lower or higher values. The Wesenheit color index ($W_{RP} - W_{K_s}$) traces temperature and molecular features in stellar spectra (see Appendix A of \citealp{lebzelter18}). 
 
We first identify the spectral types of our KELT Mira-like catalog by matching to SIMBAD using the method discussed in \S \ref{sec:catalogs}. Of the 4,132 stars in our catalog, we were able to sort 1,541 into O-rich, C-rich, or S-type. We then use our \textit{Gaia} DR2 cross-matched sample to acquire their $G_{RP}$ and $G_{BP}$ magnitudes and determined their Wesenheit indices.

\begin{figure}[t]
\centering
\includegraphics[scale = .5]{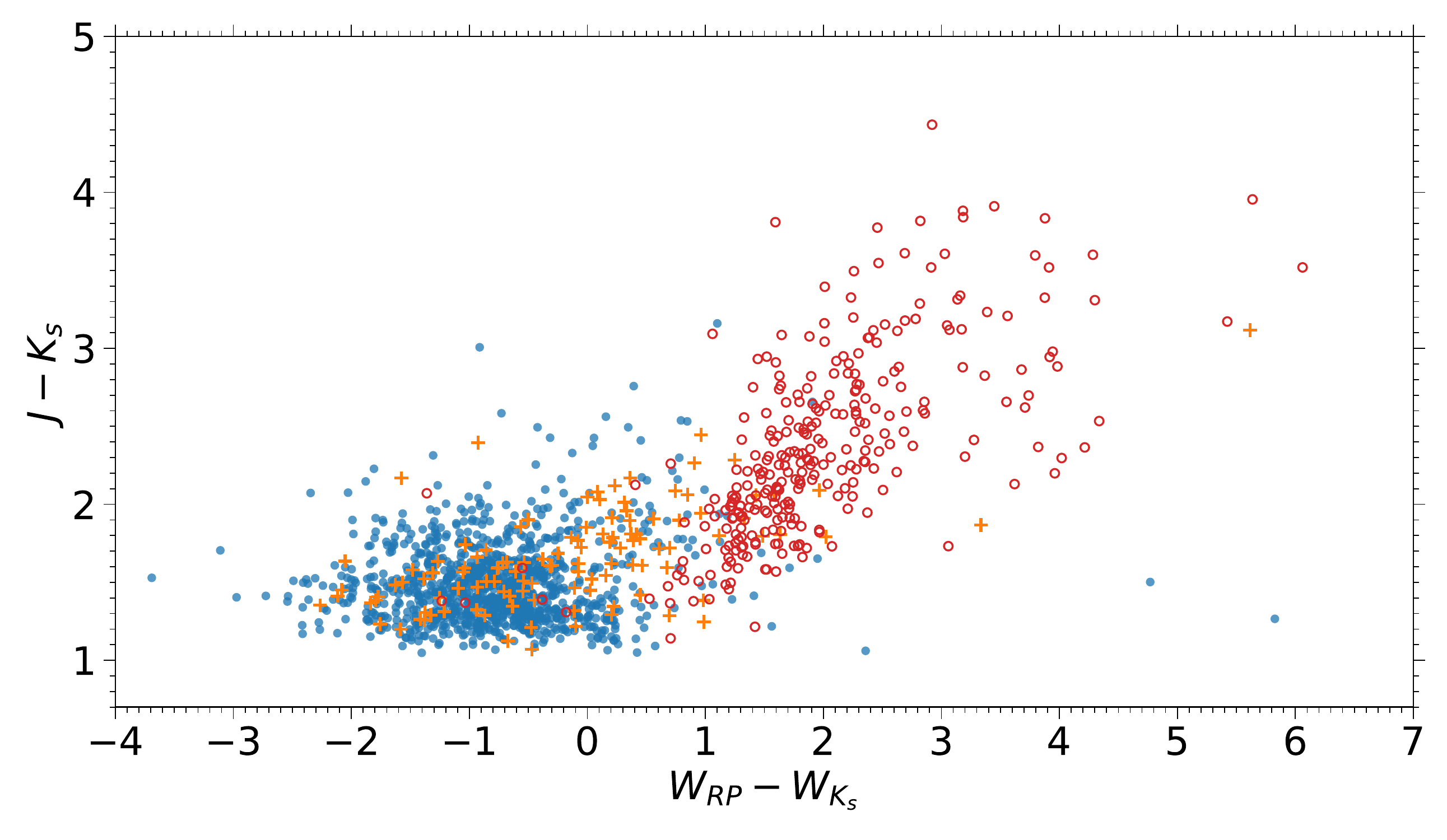}
\caption{$W_{RP} - W_{K_s}$ vs.\ $J - K_s$ for KELT Miras with known spectral types from SIMBAD. O-rich stars are shown as blue closed circles, C-rich as red open circles, and S-type in orange crosses. Two of the O-rich stars in the lower right with low $J-K_s$ and high $W_{RP} - W_{K_s}$, V CVn and RZ Sco, exhibit astrophysical masers.}
\label{fig:wesen1}
\end{figure}

The plot of ($W_{RP} - W_{K_s}$) vs.\ ($J - K_s$) for Miras with identified spectral types is shown in Figure \ref{fig:wesen1}. Known O-rich stars are shown as blue closed circles, C-rich as red open circles, and S-type as orange crosses. We can see that $W_{RP} - W_{K_s}$ is more effective than $J - K_s$ at separating O-rich from C-rich. Even without the use of absolute $K_s$ magnitudes, we are generally able to classify stars with ($W_{RP} - W_{K_s}$) \textless{} 0.8 as O-rich and those with ($W_{RP} - W_{K_s}$) \textgreater{} 0.8 as C-rich. There are three outlying O-rich Miras with low $J-K_s$ and high $W_{RP} - W_{K_s}$ colors. These three stars are V CVn, SS Oph, and RZ Sco, which respectively lie at ($W_{RP} - W_{K_s}$, $J-K_s$) = (5.8, 1.3), (4.8, 1.5), and (2.4, 1.1). Of these three, V CVn has known OH, H$_2$O, and SiO masers, and RZ Sco has a known H$_2$O maser (\citealp{kim13} and references therein).

\begin{figure}[t]
\centering 
\includegraphics[scale = .5]{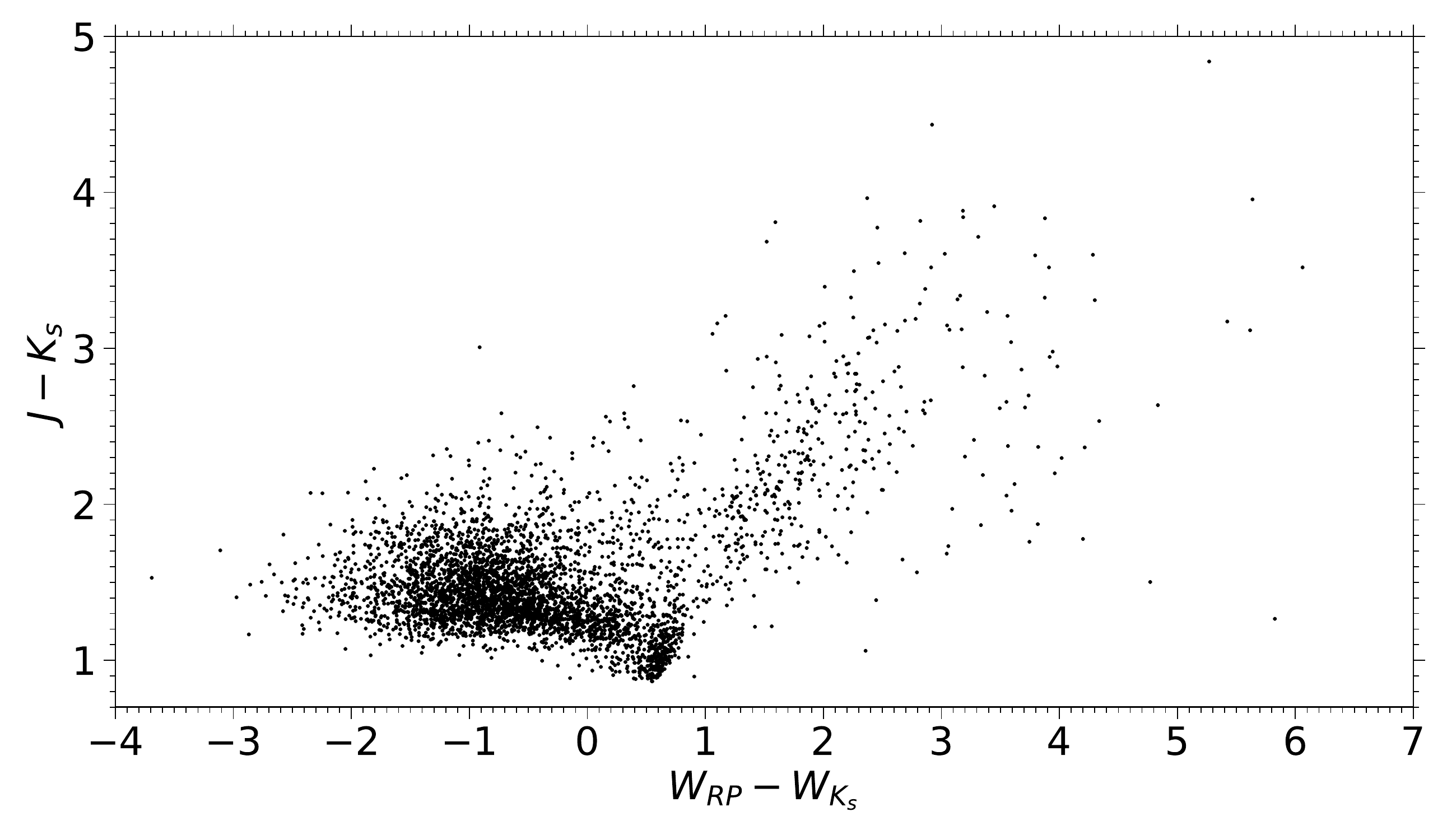}
\caption{$W_{RP} - W_{K_s}$  vs.\ $J - K_s$  for all stars in our KELT Mira-like catalog. Stars with $W_{RP} - W_{K_s}$ \textless{} 0.8 are likely O-rich and those with $W_{RP} - W_{K_s}$ \textgreater{} 0.8 are likely C-rich. The population of stars with low $J-K_s$ colors and $W_{RP} - W_{K_s}$ colors near 0.5 are likely spurious 2MASS cross-matches which are flagged in our catalog.}
\label{fig:wesen2}
\end{figure}

The plot of ($W_{RP} - W_{K_s}$) vs.\ ($J - K_s$) for our entire KELT Mira-like catalog is shown in Figure \ref{fig:wesen2}. The loci of O-rich and C-rich stars still exist but there also exists a population of approximately 250 stars with $W_{RP} - W_{K_s}$ near 0.5 and $J - K_s$ near 1.0. The colors of these stars are likely spurious as most stars in this region were found to have incorrect 2MASS identifications due to blending. These stars failed to be corrected for blending via the procedure described in \S \ref{sec:rfresults} and are flagged in our catalog (``mismatch flag" in Table 2).

\section{Summary}

We have compiled a catalog of 4,132 high-amplitude, long-period Mira-like variables using data from the KELT survey. We estimate that our catalog has a completeness of approximately 90\% for finding Miras between 8 \textless{} \textit{V} \textless{} 13, and that 70\% of our catalog objects are Miras and 30\% are SRs according to the conventional definitions. 814 objects in our catalog have not previously been classified as LPVs such as Miras, SRs, or any other class related to AGB stars.

Comparison to other catalogs reveals very good agreement between both periods, although aliasing is clearly present. The periods for most of our stars lie between 50 and 500 days, with a mean period value of 285 days. Additional peaks in the Mira period distribution, as found by V16, are not confirmed. They may not truly exist, but if they do they are masked by noise and aliasing. Future analysis on stars with longer time baselines could be done to more carefully determine the significance of these peaks. We do find that shorter periods are preferentially found at higher Galactic latitudes, consistent with \citet{whitelock94}.

Finally, we also a presented a new method to identify O-rich and C-rich stars in our Galaxy, even in the absence of absolute magnitudes, using the Wesenheit color indices discussed by \citet{lebzelter18}. This is important for future PL studies using Miras. The PL relation of O-rich Miras gives them great potential to rival Cepheids as standard candles in determining distances within the Milky Way and to nearby galaxies. Refined parallax measurements from \textit{Gaia} will improve the calibration of this PL relation which will be applicable to future distance studies.

\acknowledgments
\textit{Acknowledgments}
We would like to thank Shazrene Mohamed for her input on the early stages of this work, and R.A.A. thanks Justin Ewigleben and Alyssa Hanes for constructive criticism of the manuscript.
R.A.A. was supported by the NSF grants PHY-0849416 and PHY-1359195. 
R.A.A. would like to thank support from Lehigh University, specifically financial support through the Doctoral Travel Grant for Global Opportunities, and the Summer Research Fellowship from the College of Arts and Sciences, and support from the Department of Physics.
R.A.A. would also like to thank support from the IAU in the form of a travel grant. 
M.V.M. was supported by a Dean's Associate Professor Advancement Fellowship from Lehigh University.
P.A.W. acknowledges research funding form the South African NRF.
This research has made use of the SIMBAD database and the VizieR catalogue access tool, both operated at CDS, Strasbourg, France, and the International Variable Star Index (VSX) database, operated at AAVSO, Cambridge, Massachusetts, USA.
We have also made use of \textsc{Astropy}, a community-developed core Python package for Astronomy \citep{astropy13, astropy18}; \textsc{NumPy} \citep{vanderwalt11}; \textsc{Scikit-Learn} \citep{pedregosa12}; and \textsc{Matplotlib}, a Python library for publication quality graphics \citep{hunter07}.
This work has made use of data from the European Space Agency (ESA) mission {\it Gaia} (\url{https://www.cosmos.esa.int/gaia}), processed by the {\it Gaia} Data Processing and Analysis Consortium (DPAC,
\url{https://www.cosmos.esa.int/web/gaia/dpac/consortium}). Funding for the DPAC has been provided by national institutions, in particular the institutions participating in the {\it Gaia} Multilateral Agreement.


\begin{thebibliography}{dummy}

\bibitem[Alcock et al.(1993)]{alcock93} 
Alcock, C., Allsman, R.~A., Axelrod, T.~S., et al.\ 1993, Sky Surveys. Protostars to Protogalaxies, 291

\bibitem[Astropy Collaboration et al.(2018)]{astropy18} 
Astropy Collaboration, Price-Whelan, A.~M., Sip{\H{o}}cz, B.~M., et al.\ 2018, \aj, 156, 123

\bibitem[Astropy Collaboration et al.(2013)]{astropy13} 
Astropy Collaboration, Robitaille, T.~P., Tollerud, E.~J., et al.\ 2013, \aap, 558, A33

\bibitem[Bedding \& Zijlstra(1998)]{bedding98} 
Bedding, T.~R., \& Zijlstra, A.~A.\ 1998, \apj, 506, L47

\bibitem[Benko \& Csubry(2007)]{benko07} 
Benko, J.~M., \& Csubry, Z.\ 2007, \actaa, 57, 73

\bibitem[Blanco(1965)]{blanco65} 
Blanco, V.~M.\ 1965, in Galactic Structure, ed.\ Blaauw, A.\ \& Schmidt, M.\ (Chicago: University of Chicago Press), Vol.\ V, 241

\bibitem[Breiman(2001)]{breiman01}
Breiman, L.\ 2001,\ Machine Learning, 45, 5

\bibitem[Cadmus(2015)]{cadmus15} 
Cadmus, R.~R.\ 2015, Journal of the American Association of Variable Star Observers (JAAVSO), 43, 3

\bibitem[Catchpole et al.(2016)]{catchpole16}
 Catchpole, R.~M., Whitelock, P.~A., Feast, M.~W., et al.\ 2016, \mnras, 455, 2216

\bibitem[Chiavassa et al.(2018)]{chiavassa18} 
Chiavassa, A., Freytag, B., \& Schultheis, M.\ 2018, \aap, 617, L1

\bibitem[Cole \& Weinberg(2002)]{cole02} 
Cole, A.~A., \& Weinberg, M.~D.\ 2002, \apjl, 574, L43

\bibitem[Devor(2005)]{devor05} 
Devor, J.\ 2005, \apj, 628, 411

\bibitem[Dubath et al.(2011)]{dubath11} 
Dubath, P., Rimoldini, L., S{\"u}veges, M., et al.\ 2011, \mnras, 414, 2602

\bibitem[Dutra et al.(2002)]{dutra02} 
Dutra, C.~M., Santiago, B.~X., \& Bica, E.\ 2002, \aap, 381, 219

\bibitem[Feast et al.(1989)]{feast89}
Feast, M.~W., Glass, I.~S., Whitelock, P.~A., et al.\ 1989, \mnras, 241, 375

\bibitem[Feast et al.(1982)]{feast82} 
Feast, M.~W., Robertson, B.~S.~C., Catchpole, R.~M., et al.\ 1982, \mnras, 201, 439

\bibitem[Feast \& Whitelock(2014)]{feast14} 
Feast, M., \& Whitelock, P.~A.\ 2014, Setting the Scene for Gaia and LAMOST, 40

\bibitem[Feast et al.(2002)]{feast02}
 Feast, M., Whitelock, P., \& Menzies, J.\ 2002, \mnras, 329, L7

\bibitem[Freytag et al.(2017)]{freytag17} 
Freytag, B., Liljegren, S., \& H{\"o}fner, S.\ 2017, \aap, 600, A137

\bibitem[Gaia Collaboration et al.(2016)]{gaia16} 
Gaia Collaboration, Prusti, T., de Bruijne, J.~H.~J., et al.\ 2016, \aap, 595, A1

\bibitem[Gaia Collaboration et al.(2018)]{gaia18} 
Gaia Collaboration, Brown, A.~G.~A., Vallenari, A., et al.\ 2018, \aap, 616, A1

\bibitem[Hartman \& Bakos(2016)]{hartman16} 
Hartman, J.~D., \& Bakos, G. {\'A}.\ 2016, Astronomy and Computing, 17, 1

\bibitem[Hernitschek et al.(2016)]{hernitschek16} Hernitschek, N., Schlafly, E.~F., Sesar, B., et al.\ 2016, \apj, 817, 73

\bibitem[Huang et al.(2018)]{huang18} 
Huang, C.~D., Riess, A.~G., Hoffmann, S.~L., et al.\ 2018, \apj, 857, 67

\bibitem[Hughes \& Wood(1990)]{hughes90} 
Hughes, S.~M.~G., \& Wood, P.~R.\ 1990, \aj, 99, 784

\bibitem[Hunter(2007)]{hunter07}
Hunter, J.~D.\ 2007, Computing in Science and Engineering, 9, 90

\bibitem[Ishihara et al.(2011)]{ishihara11} 
Ishihara, D., Kaneda, H., Onaka, T., et al.\ 2011, \aap, 534, A79

\bibitem[Kerschbaum \& Hron(1992)]{kerschbaum92} 
Kerschbaum, F., \& Hron, J.\ 1992, \aap, 263, 97

\bibitem[Kim et al.(2013)]{kim13} 
Kim, J., Cho, S.-H., \& Kim, S.~J.\ 2013, \aj, 145, 22


\bibitem[Lebzelter et al.(2018)]{lebzelter18} 
Lebzelter, T., Mowlavi, N., Marigo, P., et al.\ 2018, \aap, 616, L13

\bibitem[Lomb(1976)]{lomb76} 
Lomb, N.~R.\ 1976, \apss, 39, 447

\bibitem[Matsunaga et al.(2017)]{matsunaga17} 
Matsunaga, N., Menzies, J.~W., Feast, M.~W., et al.\ 2017, \mnras, 469, 4949

\bibitem[Marigo et al.(2003)]{marigo03} 
Marigo, P., Girardi, L., \& Chiosi, C.\ 2003, \aap, 403, 225

\bibitem[Mathis(1990)]{mathis90} 
Mathis, J.~S.\ 1990, \araa, 28, 37

\bibitem[Mowlavi et al.(2018)]{mowlavi18} 
Mowlavi, N., Lecoeur-Ta{\"\i}bi, I., Lebzelter, T., et al.\ 2018, \aap, 618, A58

\bibitem[Noguchi et al.(2004)]{noguchi04} 
Noguchi, K., Aoki, W., \& Kawanomoto, S.\ 2004, \aap, 418, 67

\bibitem[Oelkers et al.(2018)]{oelkers18} 
Oelkers, R.~J., Rodriguez, J.~E., Stassun, K.~G., et al.\ 2018, \aj, 155, 39

\bibitem[Ojha et al.(2007)]{ojha07} 
Ojha, D.~K., Tej, A., Schultheis, M., et al.\ 2007, \mnras, 381, 1219

\bibitem[Pashchenko et al.(2018)]{pashch18} 
Pashchenko, I.~N., Sokolovsky, K.~V., \& Gavras, P.\ 2018, \mnras, 475, 2326

\bibitem[Payne-Gaposchkin(1951)]{payne51}
Payne-Gaposchkin, C.\ 1951, in Astrophysics, ed. Hynek, J. A. (New York, NY: McGraw-Hill), 495

\bibitem[Pedregosa et al.(2012)]{pedregosa12} 
Pedregosa, F., Varoquaux, G., Gramfort, A., et al.\ 2012, Journal of Machine Learning Research, 12, 2825

\bibitem[Pepper et al.(2012)]{pepper12} 
Pepper, J., Kuhn, R.~B., Siverd, R., et al.\ 2012, \pasp, 124, 230

\bibitem[Pepper et al.(2007)]{pepper07} 
Pepper, J., Pogge, R.~W., DePoy, D.~L., et al.\ 2007, \pasp, 119, 923

\bibitem[Pojmanski(2002)]{pojmanski02} 
Pojmanski, G.\ 2002, \actaa, 52, 397

\bibitem[Richards et al.(2011)]{richards11} 
Richards, J.~W., Starr, D.~L., Butler, N.~R., et al.\ 2011, \apj, 733, 10

\bibitem[Samus et al.(2017)]{samus17} 
Samus', N.~N., Kazarovets, E.~V., Durlevich, O.~V., et al.\ 2017, Astronomy Reports, 61, 80

\bibitem[Scargle(1982)]{scargle82} 
Scargle, J.~D.\ 1982, \apj, 263, 835

\bibitem[Schwarzenberg-Czerny(1989)]{schwarzenbergczerny89} 
Schwarzenberg-Czerny, A.\ 1989, \mnras, 241, 153

\bibitem[Skrutskie et al.(2006)]{skruskie06} 
Skrutskie, M.~F., Cutri, R.~M., Stiening, R., et al.\ 2006, \aj, 131, 1163

\bibitem[Stellingwerf(1978)]{stellingwerf78} 
Stellingwerf, R.~F.\ 1978, \apj, 224, 953

\bibitem[Soszy{\'n}ski et al.(2009)]{soszynski09} 
Soszy{\'n}ski, I., Udalski, A., Szyma{\'n}ski, M.~K., et al.\ 2009, \actaa, 59, 239

\bibitem[Soszy{\'n}ski et al.(2011)]{soszynski11} 
Soszy{\'n}ski, I., Udalski, A., Szyma{\'n}ski, M.~K., et al.\ 2011, \actaa, 61, 217

\bibitem[Soszy{\'n}ski et al.(2013)]{soszynski13} 
Soszy{\'n}ski, I., Udalski, A., Szyma{\'n}ski, M.~K., et al.\ 2013, \actaa, 63, 21

\bibitem[Stetson(1996)]{stetson96} 
Stetson, P.~B.\ 1996, \pasp, 108, 851

\bibitem[Tamuz et al.(2006)]{tamuz06} 
Tamuz, O., Mazeh, T., \& North, P.\ 2006, \mnras, 367, 1521

\bibitem[Trabucchi et al.(2017)]{trabucchi17} 
Trabucchi, M., Wood, P.~R., Montalb{\'a}n, J., et al.\ 2017, \apj, 847, 139

\bibitem[Udalski et al.(1992)]{udalski92} 
Udalski, A., Szymanski, M., Kaluzny, J., et al.\ 1992, \actaa, 42, 253

\bibitem[Uttenthaler et al.(2015)]{uttenthaler15} 
Uttenthaler, S., Blommaert, J.~A.~D.~L., Wood, P.~R., et al.\ 2015, \mnras, 451, 1750

\bibitem[VanderPlas(2018)]{vanderplas18} 
VanderPlas, J.~T.\ 2018, \apjs, 236, 16

\bibitem[van der Walt et al.(2011)]{vanderwalt11} 
van der Walt, S., Colbert, S.~C., \& Varoquaux, G.\ 2011, Computing in Science and Engineering, 13, 22

\bibitem[Vogt et al.(2016)]{vogt16} 
Vogt, N., Contreras-Quijada, A., Fuentes-Morales, I., et al.\ 2016, \apjs, 227, 6

\bibitem[Watson et al.(2006)]{watson06} 
Watson, C.~L., Henden, A.~A., \& Price, A.\ 2006, Society for Astronomical Sciences Annual Symposium, 25, 47

\bibitem[Whitelock(1996a)]{whitelock96a}
Whitelock, P. A. 1996a, in Light Curves of Variable Stars. A Pictorial Atlas, ed. Sterken, C., \& Jaschek, C. (Cambridge, United Kingdom: Cambridge University Press), 106-108

\bibitem[Whitelock(1996b)]{whitelock96b}
Whitelock, P. A. 1996b, in Light Curves of Variable Stars. A Pictorial Atlas, ed. Sterken, C., \& Jaschek, C. (Cambridge, United Kingdom: Cambridge University Press), 98-105

\bibitem[Whitelock(2012)]{whitelock12} 
Whitelock, P.~A.\ 2012, \apss, 341, 123

\bibitem[Whitelock(2013)]{whitelock13} 
Whitelock, P.~A.\ 2013, in IAU Symp.\ 289, Advancing the Physics of Cosmic Distances, ed.\ R.\ de Grijs, 209

\bibitem[Whitelock, \& Feast(2000)]{whitelock00} 
Whitelock, P., \& Feast, M.\ 2000, \mnras, 319, 759

\bibitem[Whitelock et al.(1997)]{whitelock97} 
Whitelock, P.~A., Feast, M.~W., Marang, F., et al.\ 1997, \mnras, 288, 512

\bibitem[Whitelock et al.(2006)]{whitelock06} 
Whitelock, P.~A., Feast, M.~W., Marang, F., et al.\ 2006, \mnras, 369, 751

\bibitem[Whitelock et al.(2008)]{whitelock08} 
Whitelock, P.~A., Feast, M.~W., \& Van Leeuwen, F.\ 2008, \mnras, 386, 313

\bibitem[Whitelock et al.(1994)]{whitelock94} 
Whitelock, P., Menzies, J., Feast, M., et al.\ 1994, \mnras, 267, 711

\bibitem[Wood et al.(1999)]{wood99} 
Wood, P.~R., Alcock, C., Allsman, R.~A., et al.\ 1999, in IAU Symp.\ 191, Asymptotic Giant Branch Stars, ed.\ T.\ Le Bertre, A.\ Lebre, \& C. Waelkens, 151

\bibitem[Wo{\'z}niak et al.(2004a)]{wozniak04a} 
Wo{\'z}niak, P.~R., Vestrand, W.~T., Akerlof, C.~W., et al.\ 2004a, \aj, 127, 2436

\bibitem[Wo{\'z}niak et al.(2004b)]{wozniak04b} 
Wo{\'z}niak, P.~R., Williams, S.~J., Vestrand, W.~T., et al.\ 2004b, \aj, 128, 2965

\bibitem[Yuan et al.(2017)]{yuan17} 
Yuan, W., He, S., Macri, L.~M., et al.\ 2017, \aj, 153, 170

\bibitem[Zacharias et al.(2013)]{zacharias2013} 
Zacharias, N., Finch, C.~T., Girard, T.~M., et al.\ 2013, \aj, 145, 44

\bibitem[Zechmeister \& K{\"u}rster(2009)]{zechmeister09} 
Zechmeister, M., \& K{\"u}rster, M.\ 2009, \aap, 496, 577

\end{thebibliography}
\end{document}